\documentclass[a4paper,fleqn,usenatbib]{mnras}
\usepackage[T1]{fontenc}
\usepackage{ae,aecompl}

\usepackage[pdftex]{graphicx}	% Including figure files
\usepackage{here}
\usepackage{amsmath}	% Advanced maths commands
\usepackage{amssymb}	% Extra maths symbols

\usepackage{lscape}
\usepackage{graphicx}
\usepackage{color}
\usepackage{bm}
\usepackage{comment}
\usepackage{newtxtext,newtxmath}
\title[MHD effect on the first star formation]{Magnetohydrodynamic effect on first star formation: prestellar core collapse and protostar formation}

\author[K. Sadanari et al.]{
Kenji Eric Sadanari,$^{1}$\thanks{E-mail: k.sadanari@astr.tohoku.ac.jp}
Kazuyuki Omukai,$^{1}$
Kazuyuki Sugimura,$^{1,2}$
Tomoaki Matsumoto,$^{3}$
 and Kengo Tomida$^{1}$
\\
$^{1}$Astronomical Institute, Tohoku University, Aoba, Sendai, Miyagi 980-8578, Japan\\
$^{2}$Department of Astronomy, University of Maryland, College Park, MD 20740, USA\\
$^{3}$Faculty of Sustainability Studies, Hosei University, Fujimi, Chiyoda, Tokyo 102-8160, Japan\\
}

\date{Accepted XXX. Received YYY; in original form ZZZ}

\pubyear{2021}

\begin{document}
\label{firstpage}
\pagerange{\pageref{firstpage}--\pageref{lastpage}}
\maketitle

% Abstract of the paper
\begin{abstract}
Recent theoretical studies have suggested that a magnetic field may play a crucial role in the first star formation in the universe. 
However, the influence of the magnetic field on the first star formation has yet to be understood well.
In this study, 
we perform three-dimensional magnetohydrodynamic simulations taking into account all the relevant cooling processes and non-equilibrium chemical reactions up to the protostar density,
in order to study the collapse of magnetized primordial gas cores with self-consistent thermal evolution. 
Our results show that the thermal evolution of the central core is hardly affected by a magnetic field, 
because magnetic forces do not prevent the contraction along the fields lines.
We also find that the magnetic braking extracts the angular momentum from the core and suppresses fragmentation depending on the initial strength of the magnetic field. 
The angular momentum transport by the magnetic outflows is less effective than that by the magnetic braking because the outflows are launched only in a late phase of the collapse.
Our results indicate that the magnetic effects become important for the field strength $B> 10^{-8}(n_{\rm H}/1\ \rm cm^{-3})^{2/3}\ \rm G$, where $n_{\rm H}$ is the number density, during the collapse phase.
Finally, we compare our results with simulations using a barotropic approximation and confirm that this approximation is reasonable at least for the collapse phase.
Nevertheless, self-consistent treatment of the thermal and chemical processes is essential for extending simulations to the accretion phase, in which radiative feedback by protostars plays a crucial role.
\end{abstract}

% Select between one and six entries from the list of approved keywords.
% Don't make up new ones.
\begin{keywords}
stars:formation, stars: Population$\rm I\hspace{-.1em}I\hspace{-.1em}I$, stars:magnetic field
%accretion, accretion discs -- cosmology: theory -- dark ages, reionization, first stars
\end{keywords}

%%%%%%%%%%%%%%%%%%%%%%%%%%%%%%%%%%%%%%%%%%%%%%%%%%

%%%%%%%%%%%%%%%%% BODY OF PAPER %%%%%%%%%%%%%%%%%%

\section{Introduction}
The first stars, also known as the Pop III stars, 
are thought to have formed around the redshift $z \sim 20-30$ in minihalos of $M\sim10^{6}\ M_{\odot}$ 
(\citealp{Couchman_Rees1986}; \citealp{Yoshida2003}; \citealp{Greif2015}).  
They play crucial roles in the subsequent structure formation in the universe.
For example:
ionizing photons emitted from these stars start reionizing the intergalactic medium; 
their supernovae spread heavy elements into the interstellar medium, which drive the chemical evolution of the universe (e.g., \citealp{Ciardi_Ferrara2005}).
The first stars are formed from the primordial-gas, which lacks metals and dust grains. 
Only available coolant in the low temperature ($<10^{4}\ \rm K$) is molecular hydrogen $\rm H_{\rm 2}$, 
resulting in rather high temperature of several hundred K during the prestellar collapse and thus high accretion rate $\sim 10^{-3}\  M_{\sun}/{\rm yr}$ onto the formed protostar (\citealp{Stahler_Palla1986}; \citealp{Omukai_Nishi1998}).
The high accretion rate causes formation of massive stars with $10-1000\ M_{\odot}$ 
(\citealp{Abel2002}; \citealp{Bromm2002}; \citealp{Omukai_Palla2003}; \citealp{Yoshida2008}; \citealp{McKee_Tan2008}; \citealp{Hosokawa2016}; \citealp{Stacy2016}),
and their initial mass function (IMF) is expected to be top heavy (\citealp{Hirano2014}; \citealp{Susa2014}).\par
Previous works have demonstrated by three-dimensional (3D) simulations that first stars tend to be in binary (or multiple) systems as a result of prestellar fragmentation during the collapse or circumstellar disc fragmentation in the accretion phase 
(e.g., \citealp{Machida2008b}; \citealp{Clark2011}; \citealp{Smith2011}; \citealp{Stacy2016}; \citealp{Susa2019}; \citealp{Chon_Hosokawa2019}; \citealp{Kimura2020}).
\citet{Sugimura2020} have performed 3D radiation hydrodynamics simulations starting from the onset of the gravitational collapse to the end of accretion, 
and have found the formation of a similar-mass binary system consisting of stars with $56$ and $66\ M_{\odot}$ at a separation of $2\times 10^{4}\ \rm au$.
Such a massive binary is expected to evolve into a black hole (BH) binary system.
If close enough, a BH binary merges within the age of the universe and can be sources of observed gravitational wave (GW) events.
Recent population synthesis calculations (e.g., \citealp{Kinugawa2014}, \citeyear{Kinugawa2016}) suggest that $\sim 30\ M_{\odot}$ BH binaries observed by GWs are originated from the first star.\par
Those 3D simulations, however, have not included effects of the magnetic field. 
In nearby star forming regions, magnetic fields of several $\rm \mu G$ are known to exist.
Such strong fields in the present-day universe are believed to be originated from amplification of weak seed fields by galactic dynamo processes (e.g., \citealp{Brandenburg_Subramanian2005}; \citealp{Pakmor2017}).
From a theoretical point of view, seed fields of $10^{-20}- 10^{-10}\ \rm{G}$
can be generated from astrophysical mechanisms involving the battery effect (\citealp{Biermann1950}) due to supernovae (\citealp{Hanayama2005}), galaxy formation (\citealp{Kulsrud1997}), reionization (\citealp{Gnedin2000}; \citealp{Attia2021}), and radiation pressure from the radiation sources (\citealp{Langer2003}; \citealp{Doi_Susa2011}).
Seed fields could also be generated in the very early universe during the electroweak and QCD phase transitions 
(e.g., \citealp{Baym1996}; \citealp{Quashnock1989}; \citealp{Banerjee_Jedamzik2004}; \citealp{Durrer_Neronov2013}; \citealp{Subramanian2016}), 
although the generated field strength rely highly on uncertainties in the adopted models.
In addition, \citet{Ichiki2006} show that the cosmological fluctuations can generate seed magnetic fields of about $10^{-18}\ \rm{G}$ at $1\ \rm{Mpc}$ scale during the recombination epoch. 
It is thus likely that seed magnetic fields already existed before the formation of first stars.
Such weak fields, however, have little effects on the first star formation.
In order to drive a protostellar jet during collapse phase, 
the host cloud with the density of about $1\ \rm cm^{-3}$ needs to have a field stronger than $10^{-11}\ \rm{G}$ (\citealp{Machida2008}). 
Therefore magnetic fields play a significant role in the first star formation only if the seed field can be sufficiently amplified.  
\citet{Schober2012} have investigated the amplification of magnetic fields by small-scale dynamo during the collapse of a turbulent first-star-forming cloud by one-zone calculation and found that the field increases exponentially immediately after the onset of the collapse 
and reaches the level as strong as $\sim 10^{-6}\ \rm G$.
The small-scale dynamo is also shown to be effective in amplifying the magnetic field by 3D hydrodynamical simulations 
(e.g., \citealp{Sur2010}; \citealp{Federrath2011b}; \citealp{Turk2012}; \citealp{Sharda2020b}).
With amplification of small initial seeds, 
magnetic fields may have substantial effects on the first star formation.\par
3D magnetohydrodynamic (MHD) simulations for the first star formation have already been performed (e.g., \citealp{Machida2008}; \citealp{Sharda2020a}).
\citet{Machida2008} have calculated prestellar collapse of primordial-gas clouds until the protostar formation at $10^{21} \ \rm cm^{-3}$ with different sets of the initial rotation and magnetic energy by 3D MHD simulations, and investigated the condition for fragmentation and outflow launching.
They found that if the initial magnetic field has higher energy than that of the cloud rotation, the angular momentum of the cloud is transported by the magnetic braking and suppresses the formation of a rotation-supported disc and its fragmentation.
They also showed that MHD winds are driven in models in which the fragmentation does not occur. 
In such cases, MHD winds may play an important role in setting the star formation efficiency in cloud cores, as in the present-day star formation.
Recently, \citet{Sharda2020a} showed that even weaker magnetic fields ($10^{-15}\ \rm G$) can affect the IMF of first stars by extending simulations to the accretion phase.\par
However, \citet{Machida2008} used the barotropic approximation for the gas equation of state, where the temperature is given as a function of density, 
based on the result of the one-zone calculation (\citealp{Omukai2005}), without solving the thermal evolution of gas consistently. 
Effects such as shock heating and back-reaction of strong magnetic fields on the temperature evolution (\citealp{Nakauchi2019}) cannot be properly incorporated in barotropic simulations. 
It is not clear to what extent these effects influence the dynamics, such as the cloud fragmentation by changing the temperature evolution.\par
In this work, we reveal magnetic effects on the collapse of primordial-gas cloud cores by solving the thermal evolution and gas dynamics consistently.
To this end, for the first time, 
we perform 3D MHD simulations of collapse phase taking into account all the relevant cooling processes and non-equilibrium chemical reaction up to the protostar density.
We also perform barotropic simulations for comparison and discuss the validity of this approximation.\par
This paper is organized as follows.
In Section \ref{sec2}, we describe the method and setup of our simulations.
We present the results in Section \ref{sec3}.
In section \ref{sec3.1}, we first focus on the evolution of non-rotating clouds to see magnetic field effects on the central temperature evolution, and in section \ref{sec3.2}, we present the results for magnetized rotating clouds to see effects of the magnetic braking and outflows.
In section \ref{sec3.3}, we discuss the fragmentation and outflow-launching conditions for the rotational and magnetic energy of the initial cloud.
Finally, in section \ref{sec3.4}, we compare our results with those obtained with the barotropic approximation and discuss its validity. 
In section \ref{sec4}, we summarize our findings and discuss magnetic field effects on the first star formation.
%------------------------------
 \begin{table}
 \caption{Chemical reactions}
 \label{table1}
 \centering
  \begin{tabular}{cl}
   \hline \hline
   Number  &  Reaction\\
   \hline 
   H1   &  ${\rm H^{+} + e \rightleftharpoons H + \gamma}$ \\
   H2    &  ${\rm H + e \rightleftharpoons H^{-} + \gamma}$ \\
   H3    &  ${\rm H^{-} + H \rightleftharpoons H_{2} + e}$ \\
   H4    &  ${\rm H + H^{+} \rightleftharpoons H_{2}^{+} + \gamma}$ \\
   H5    &  ${\rm H_{2}^{+} + H \rightleftharpoons H_{2} + H^{+}}$ \\
   H6    &  ${\rm 3H  \rightleftharpoons H_{2} + H}$ \\
   H7    &  ${\rm 2H_{2} \rightleftharpoons 2H + H_{2}}$ \\
   D1    &  ${\rm D + H^{+} \rightleftharpoons D^{+} + H}$ \\
   D2    &  ${\rm D + H_{2} \rightleftharpoons H + HD}$ \\
   D3    &  ${\rm D^{+} + H_{2} \rightleftharpoons H^{+} + HD}$ \\
   \hline
  \end{tabular}
\end{table}
\label{Sec:introduction}
%-----------------------------------------
%%%%%%%%%%%%%%%%%%%%%%%%%%%%%%%%%%%%%%%%%%%%%%%%%%%%%%%%%%%%%%%%%%%%%%%
%%%%%%%%%%%%%%%%%%%%%%%%%%%%%%%%%%%%%%%%%%%%%%%%%%%%%%%%%%%%%%%%%%%%%%%
\section{Numerical Method}\label{sec2}
We follow collapsing cloud evolution by using a magnetohydrodynamics code SFUMATO (\citealp{Matsumoto2007}; \citealp{Matsumoto2015}) with adaptive mesh refinement (AMR) and self-gravity.
The MHD part is solved by the HLLD approximation Riemann solver (\citealp{Miyoshi_Kusano2005}) with a mixed divergence cleaning method (\citealp{Dedner2002}), 
and the self-gravity part is solved by a multigrid method.
These two schemes have a second-order accuracy in space and time.
We modify SFUMATO by introducing a module solving the chemical and thermal evolution of primordial gas,
which is used in a radiative hydrodynamics version of the code, SFUMATO-RT (\citealp{Sugimura2020}). 
In addition, since the thermal processes in the SFUMATO-RT are only relevant in the density range $n_{\rm H} < 10^{13}\ \rm cm^{-3}$, 
we add the chemical and cooling processes that are important in a higher density region to solve until the protostar formation.\par 
The governing equations are as follows:
%------------------
 the mass conservation, 
\begin{equation} \label{eq1}
\frac{\partial \rho}{\partial t} + \nabla  \cdot \left( \rho \bm{v} \right)=0,
\end{equation}
%---------------
the equation of motion, 
\begin{equation} \label{eq2}
\rho \frac{\partial \bm{v}}{\partial t} + \rho \left( \bm{v} \cdot \nabla \right) \bm{v} = - \nabla p- \frac{1}{4 \pi} \bm{B} \times \left( \nabla \times \bm{B}\right) - \rho \nabla \phi,
\end{equation}
%------------------
the gas energy equation, 
\begin{equation} \label{eq3}
\frac{\partial e}{\partial t} + \nabla \cdot \left[  \left(  e + p + \frac{|\bm{B}|^{2}}{8\pi}   \right) \bm{v} -  \frac{1}{4\pi} \bm{B} \left(  \bm{v} \cdot \bm{B}\right) \right] = -\rho \bm{v} \cdot \nabla \phi -\Lambda,
\end{equation}
%----------
the induction equation of ideal MHD, 
\begin{equation} \label{eq4}
\frac{\partial \bm{B}}{\partial t} = \nabla \times \left(  \bm{v}\times \bm{B}  \right),
\end{equation}
%--------------
the solenoidal constraint, 
\begin{equation} \label{eq5}
\nabla \cdot \bm{B} = 0,
\end{equation}
%-----------------
and the Poisson equation of the gravity, 
\begin{equation} \label{eq6}
\nabla ^{2} \phi = 4 \pi G \rho,
\end{equation}
%-----------------
where $\rho$, $p$, $\bm{v}$, $\bm{B}$, $\phi$, $e$, $\Lambda$ are the gas density, gas pressure, gas velocity, magnetic field, gravitational potential, total gas energy per unit volume and net cooling rate per unit volume, respectively.
The energy density $e$ is given by
%-------------------------------
\begin{equation}
e = \frac{1}{2}\rho |\bm{v}|^{2} +\frac{p}{\gamma-1}+\frac{1}{8\pi}|\bm{B}|^{2}, 
\end{equation}
%----------------------------------
where $\gamma$ is the adiabatic index, which depends on the chemical composition and gas temperature (e.g., \citealp{Omukai_Nishi1998}).\par
We initially take a cube with $L_{\rm{box}}=4\times10^{6}\ \rm{au}$ on one side as the computational domain and set base grids with $N_{\rm{base}}=64$ cells in each direction.
When the width of the cell $h(l)$ at the grid level $l$ exceeds 1/16 of the local Jeans length, the cell is refined with the grid level raised to $l+1$.
The cell width is halved at each refinement of the grid: $h(l+1)=h(l)/2$.
We set the maximum grid level $l_{\rm{max}}=25$, and thus the minimum width of the cell in this simulation is $\Delta x_{\rm{min}}=L_{\rm{box}}/N_{\rm{base}}\times2^{-l_{\rm{max}}} \simeq 2\times10^{-3}\ \rm{au}$.
In our simulation, 
we calculate the cloud collapse until the protostar formation at $n_{\rm H} \simeq 10^{20}\ \rm cm^{-3}$ without using sink particles. 
\subsection{Thermal and chemical processes}
The net cooling rate $\Lambda$ in Equation (\ref{eq3}) is given as
%-----------------
\begin{equation} \label{eq7_4}
\Lambda = \Lambda_{\rm{line}} + \Lambda_{\rm{cont}} + \Lambda_{\rm{chem}},
\end{equation}
%----------------------
where $\Lambda_{\rm{line}}$, $\Lambda_{\rm{cont}}$, $\Lambda_{\rm{chem}}$ are the line cooling, continuum cooling and chemical cooling/heating, respectively.
The line cooling  consists of the $\rm{H_{2}}$ and HD cooling: $\Lambda_{\rm{line}}=\Lambda_{\rm{H_{2}}}+\Lambda_{\rm{HD}}$.
The H$_2$ and HD line cooling rate is calculated as
%------------------------
\begin{equation} \label{eq8_3}
\Lambda_{\rm M} = \bar{\beta}_{\rm esc, M} \rm{e}^{-\tau} \Lambda_{\rm M,thin},
\end{equation}
%----------------------
where $\rm M$ represents the molecular species H$_2$ or HD, and the optically-thin cooling rate $\Lambda_{\rm{M, thin}}$ is 
taken from the fitting function of \citet{Glover2015} and \citet{Lipovka2005} for H$_{2}$ and HD, respectively.
The photon trapping effect is included by using $\bar{\beta}_{\rm{esc}}$ and $\rm{e}^{-\tau}$ for the line and continuum absorption, respectively.
For the line averaged escape probability $\bar{\beta}_{\rm{esc}}$, 
we use the fitting function given in \citet{Fukushima2018}, which 
depends on the column density $N_{\rm M}$ of molecular species $\rm M$ and the gas temperature $T$. 
The column density $N_{\rm M}$ is estimated from the local Jeans length as $N_{\rm M}=n_{\rm M} \lambda_{\rm{J}}$,
where, $n_{\rm M} $ is the local number density of the species $M$.
The effective continuum optical depth $\tau$ is given by the geometrical average of the Planck and Rosseland mean optical depths, $\tau=\sqrt{\tau_{\rm{P}} \tau_{\rm{R}}}$, 
where the Planck (Rosseland) mean optical depths $\tau_{\rm P(R)}$ are again estimated from those of the local Jeans length: $\tau_{\rm{P(R)}}= \kappa_{\rm{P(R)}} \rho \lambda_{\rm{J}}$. 
The Planck (Rosseland) mean opacity $\kappa_{\rm{P(R)}}$ are 
taken from \citet{Matsukoba2019}.
For the continuum cooling, we consider
the H free-bound emission, $\rm{H}^{-}$ free-bound emission, $\rm{H}^{-}$ free-free emission, $\rm{H}$ free-free emission, $\rm{H_{2}}$-$\rm{H_{2}}$ collision-induced emission (CIE) and $\rm{H_{2}}$-$\rm{He}$ CIE. 
The photon trapping effect is taken into account following \citet{Tanaka_Omukai2014}: the continuum cooling rate $\Lambda_{\rm cont}$ is given as
%-----------------
\begin{equation} \label{eq9_1}
\Lambda_{\rm{cont}} = \frac{\Lambda_{\rm{cont,thin}}}{1+\tau_{\rm{P}}+\frac{3}{4}\tau_{\rm{R}} \tau_{\rm{P}}},
\end{equation}
%--------------------------
with the fitting function of optically thin case $\Lambda_{\rm {cont,thin}}$, which depends on the gas density and temperature from \citet{Matsukoba2019}.
The net chemical cooling rate $\Lambda_{\rm{chem}}$ is calculated by considering the $\rm{H}$ ionization/recombination and $\rm{H_{2}}$ dissociation/formation: 
%-----------------
\begin{equation} \label{eq10}
\Lambda_{\rm{chem}} = \left( \chi_{\rm{H}} \frac{\rm{d}y(\rm{H^{+})}}{\rm{d}t} - \chi_{\rm{H_{2}}} \frac{\rm{d}y(\rm{H_{2}})}{\rm{d}t}  \right)n_{\rm{H}},
\end{equation}
%-----------------
where $\chi_{\rm{H}}=13.6\ \rm{ev}$ and $\chi_{\rm{H_{2}}}=4.48\ \rm{ev}$ are the binding energies, $y(i)$ is the chemical fraction of species $i$ relative to the hydrogen nuclei, i.e., $y(i)=n(i)/n_{\rm{H}}$, with $n(i)$ and $n_{\rm H}$ the number density of species $i$ and hydrogen nuclei, respectively.\par
We consider 20 chemical reactions among the nine species: $\rm{H}$, $\rm{H^{+}}$, $\rm{H_{2}}$, $\rm{e}$, $\rm{H^{-}}$, $\rm{H_{2}^{+}}$, $\rm{D}$, $\rm{HD}$, $\rm{D^{+}}$. 
We assume all the $\rm{He}$ is neutral with $y(\rm{He})=9.722\times10^{-2}$. 
Table \ref{table1} shows the 20 chemical reactions selected from the minimal chemical reactions presented in \citet{Nakauchi2019}.
The reaction rate coefficients are also taken from \citet{Nakauchi2019}.
We solve the non-equilibrium chemistry at density $n_{\rm H}<10^{18}\ \rm cm^{-3}$, 
but at higher density,
set the chemical abundances to the equilibrium values 
since the chemical reaction time scale is at least an order of magnitude smaller than free-fall time.
% -----------
\begin{table}
 \caption{Model parameters}
 \label{table2}
 \centering
  \begin{tabular}{lccccc}
   \hline \hline
   Model  &  $ E_{\rm{r}}/|E_{\rm g}|$  &  $E_{\rm{m}}/|E_{\rm g}|$ & $\Omega_{0}\ \rm[s^{-1}]$ & $B_{0}\ \rm[G]$ & $\mu_{0} $\\
 
   \hline 
   1..........    &  0             &  0                            &  0                                  &     0                               &   $\infty$ \\
   2 ........  &   0             & $2\times10^{-5}$     &  0                                  &    $9.6\times10^{-8}$     &  270  \\ 
   3 ........   &   0             & $2\times10^{-3}$     &  0                                  &    $9.6\times10^{-7}$     &  27    \\
   4 ........   &   0             &  $2\times10^{-1}$    &  0                                 &     $9.6\times10^{-6}$     &  2.7\\
   5 ........   &   0             &  $9.5\times10^{-1}$ &  0                                 &     $2.1\times10^{-5}$     &   1.2\\
   6 ........  & $10^{-6}$  &  0                            &  $2.9\times10^{-17}$    &   0                            &   $\infty$ \\
   7 ........   & $10^{-6}$  &  $2\times 10^{-5}$   &  $2.9\times10^{-17}$    &  $9.6\times10^{-8}$    &  270  \\
   8 ........   &  $10^{-6}$ & $2\times10^{-3}$     &  $2.9\times10^{-17}$    &  $9.6\times10^{-7}$      &  27    \\
   9 ........   &  $10^{-6}$ &  $2\times10^{-1}$    &  $2.9\times10^{-17}$    &  $9.6\times10^{-6}$     &  2.7\\ 
   10 ..... & $10^{-4}$  &  0                            & $2.9\times10^{-16}$     &  0                            &   $\infty$ \\
   11 ..... & $10^{-4}$  &  $2\times 10^{-5}$   & $2.9\times10^{-16}$     &  $9.6\times10^{-8}$    &  270  \\
   12 ..... &  $10^{-4}$ & $2\times10^{-3}$     & $2.9\times10^{-16}$    &  $9.6\times10^{-7}$      &  27    \\
   13 ..... &  $10^{-4}$ &  $2\times10^{-1}$    & $2.9\times10^{-16}$    &  $9.6\times10^{-6}$      &  2.7\\
   14 ..... & $10^{-2}$  &  0                            & $2.9\times10^{-15}$    &  0                             &   $\infty$ \\
   15 ..... & $10^{-2}$  &  $2\times 10^{-5}$   & $2.9\times10^{-15}$    &  $9.6\times10^{-8}$     &  270  \\
   16 ..... &  $10^{-2}$ & $2\times10^{-3}$     & $2.9\times10^{-15}$    &  $9.6\times10^{-7}$     &  27    \\
   17 ..... &  $10^{-2}$ &  $2\times10^{-1}$    & $2.9\times10^{-15}$    &  $9.6\times10^{-6}$     &  2.7\\
   \hline
  \end{tabular}
\end{table}
%----------------------------
\begin{figure*}
\centering
\includegraphics[trim=0 0 0 0, scale = 0.35, clip]{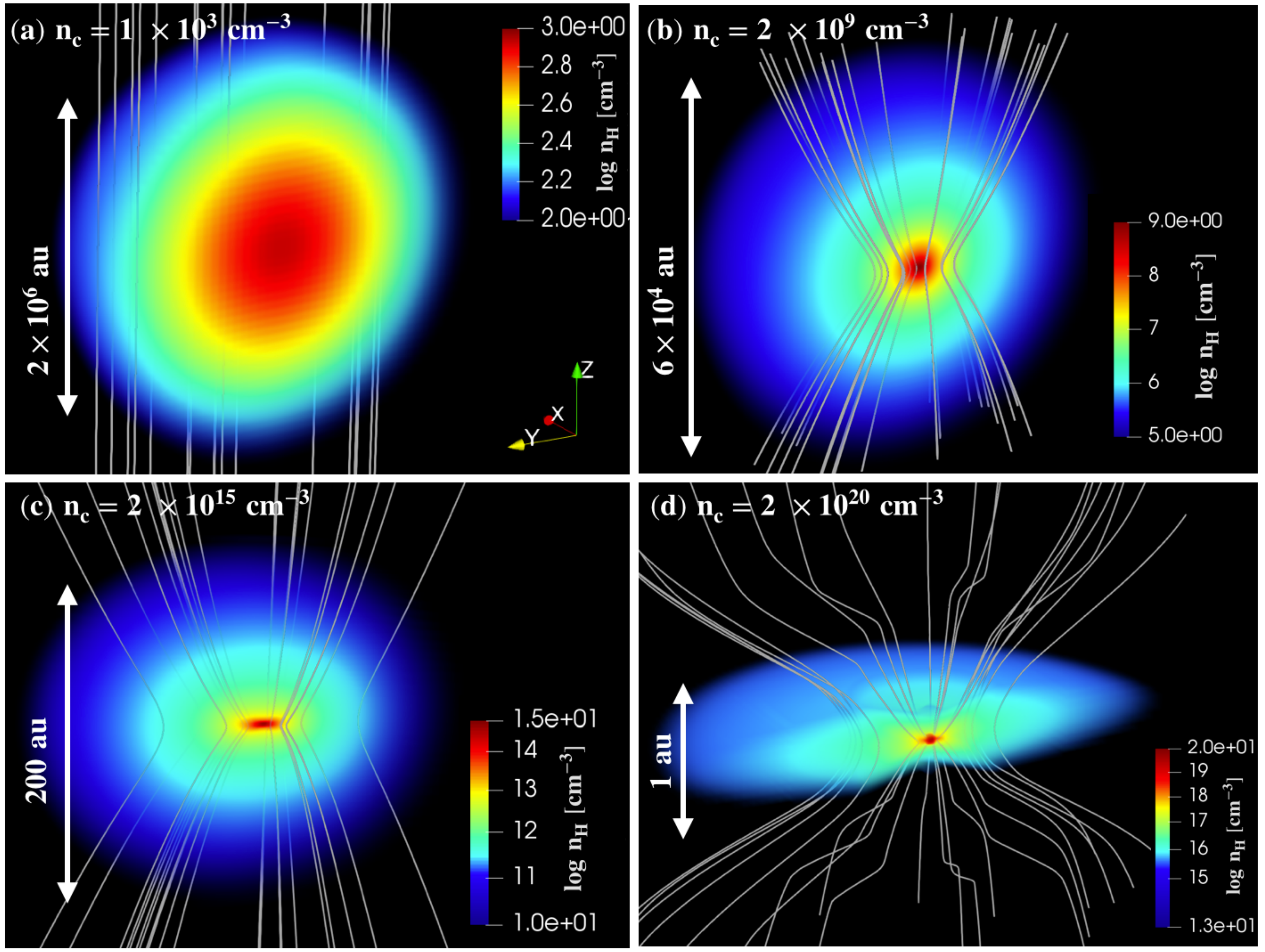}
\caption{
3D volume rendering of the density field with the magnetic field lines for the non-rotating weak magnetic field model ($E_{\rm m}/|E_{\rm g}| = 2\times10^{-3}$). 
To make the central regions visible, only the density data on the rear side of the $y$-$z$ plane ($x=0$) is used for rendering.
Each panel shows the cloud structure at four stages: the central density $n_{\rm c}=$ (a)$ 1\times10^3 \ \rm cm^{-3}$(initial state), (b) $2\times10^9 \ \rm cm^{-3}$, (c) $2\times 10^{15}\ \rm cm^{-3},$ and (d) $2\times10^{20}\ \rm cm^{-3}$ (protostar formation).
}
\label{fig1}
\end{figure*}
% --------------------
\subsection{Initial conditions}
We set the initial conditions following \citet{Machida2008}. 
As the initial density profile of cloud cores, we use the critical Bonnor-Ebert sphere (\citealp{Ebert1955}; \citealp{Bonnor1956}) $\rho_{\rm{BE}}(r)$, which is the isothermal sphere in hydrostatic equilibrium with external pressure.
In order to promote a core collapsing, we increase the density profile by a factor $f=1.4$.
We also impose density fluctuations to induce the fragmentation during collapse:
%-------------------------
\begin{equation} \label{eq9}
\rho(r,\varphi)= \left\{
                   \begin{array}{ll}
                   f \bigl[ 1+\delta \rho(r, \varphi) \bigr]\rho_{\rm{BE}}(r)   \ \ \ {\rm for}\ \ \sl{r<\ R_{0}}, \\
                   \\
                   f \bigl[ 1+\delta \rho(r, \varphi) \bigr] \rho_{\rm{BE}}(R_{0}) \ {\rm for}\ \ \sl{r \geq \ R_{0}},
                   \end{array}
                   \right.
 \end{equation}
 %-----------------------
where the initial cloud radius $R_{0}=1.1\times10^{6}\ \rm{au}$.
In this work, we assume a small m=2-mode density perturbation $\delta \rho$, such as,
%----------------------------------------
\begin{equation} \label{eq8_2}
\delta \rho(r, \varphi)= A_{\rm{\varphi}}(r/R_{0})^2 \cos{2\varphi},
\end{equation}
%----------------------------------------------
where the amplitude $A_{\rm{\varphi}}=0.01$. 
Note that the initial density profile is uniquely determined by the central density and the isothermal temperature, for which we adopt $\rho_{\rm{BE}}(0)=2.1\times10^{-21}\ \rm{g\ cm^{-3}}$ (number density $n_{\rm{BE}}(0)=10^{3}\ \rm{cm^{-3}}$) and $T_{\rm{iso}}=198\ \rm{K}$.
This temperature is taken from the value at the density $n_{\rm{BE}}(0)f=1.4\times10^{3}\ \rm{cm^{-3}}$ obtained by the one-zone model of the cloud collapse with the same thermal and chemical processes.  
The total mass enclosed inside the initial radius $R_{0}$ is $M_{\rm{c}}=5.5\times10^{3}\  M_{\odot}$ and 
the ratio of the gravitational energy to the thermal energy $\alpha=E_{\rm th}/|E_{\rm g}|=0.6$.\par
We assume that the cloud is initially rigidly rotating around the $z$-axis with an angular velocity $\Omega_{0}$ and pierced by a uniform magnetic field $B_{0}$ parallel to the rotation axis.
As the boundary conditions, we also assume that both the magnetic fields and ambient gas outside the initial Bonner-Ebert sphere rotate at the initial angular velocity of the cloud (\citealp{Matsumoto_Tomisaka2004}).
The initial state is characterized with two parameters: the ratio of the rotational energy to the gravitational energy $E_{\rm{r}}/|E_{\rm g}|$ and that of the magnetic energy to the gravitational energy $E_{\rm{m}}/|E_{\rm g}|$.
Table \ref{table2} summarizes the parameters of 17 models examined in this study.
We also estimate the initial mass-to-flux ratio,
%---------------------------------
\begin{equation} \label{eq8_1}
\left(\frac{M}{\Phi} \right)= \frac{M_{\rm c}}{\pi R_{0}^2 B_{0}},
\end{equation}
%----------------------------------------
where $\Phi$ denotes the magnetic flux.
The dimensionless parameter $\mu_{0}$ in Table \ref{table2} indicates the mass-to-flux ratio normalized by the critical value $(M/\Phi)_{\rm cr}$ for a cloud with uniform density (\citealp{Mouschovias_Spitzer1976})
%---------------------------------------
\begin{equation} \label{eq8_4}
\left(\frac{M}{\Phi} \right)_{\rm cr}= \frac{0.53}{3\pi}\left( \frac{5}{G}\right)^{\frac{1}{2}}.
\end{equation}
%----------------------------------------------
\label{Sec:method}
%%%%%%%%%%%%%%%%%%%%%%%%%%%%%%%%%%%%%%%%%%%%%%%%%%%%%%%%%%%%%%%%%%%%%%%%%
%%%%%%%%%%%%%%%%%%%%%%%%%%%%%%%%%%%%%%%%%%%%%%%%%%%%%%%%%%%%%%%%%%%%%%%
%----------------------------------------------------------------
\begin{figure}
\centering
\includegraphics[trim=0 35 0 65, scale = 1.00, clip]{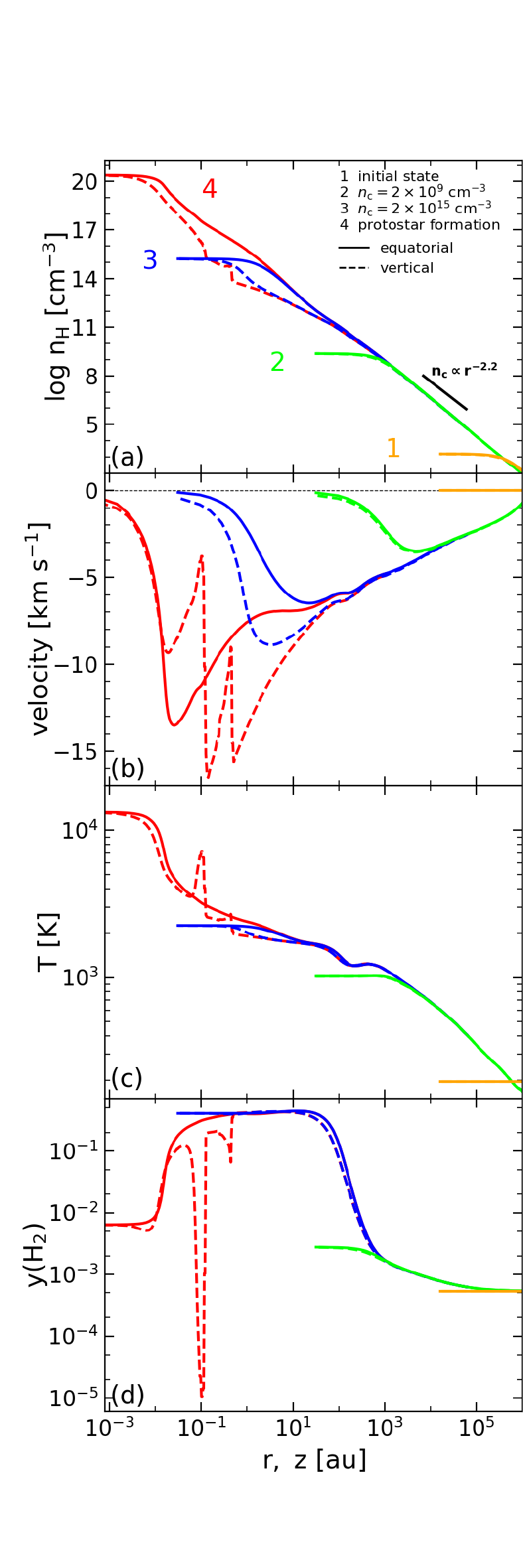}
\caption{
1D gas profiles for the weak magnetic field model ($E_{\rm m}/|E_{\rm g}| = 2\times10^{-3}$) at the same stages as in Fig. \ref{fig1}, where physical quantities on the equatorial plane are azimuthally averaged.
We provide the profiles of the (a) density, (b) velocity, (c) temperature, and (d) $\rm H_{2}$ abundance in the equatorial (solid) and vertical (dashed) directions.
In panel (b), the solid and dashed lines represent the equatorial and vertica velocities $v_{ r}$ and $v_{z}$, respectively.
The yellow, green, blue, and red colors indicate the four stages of Fig. \ref{fig1}(a-d), respectively.
}
\label{fig2}
\end{figure}
%-----------------------------------------------------------------
\begin{figure}
\centering
\includegraphics[trim=0 5 0 32, scale = 1.05, clip]{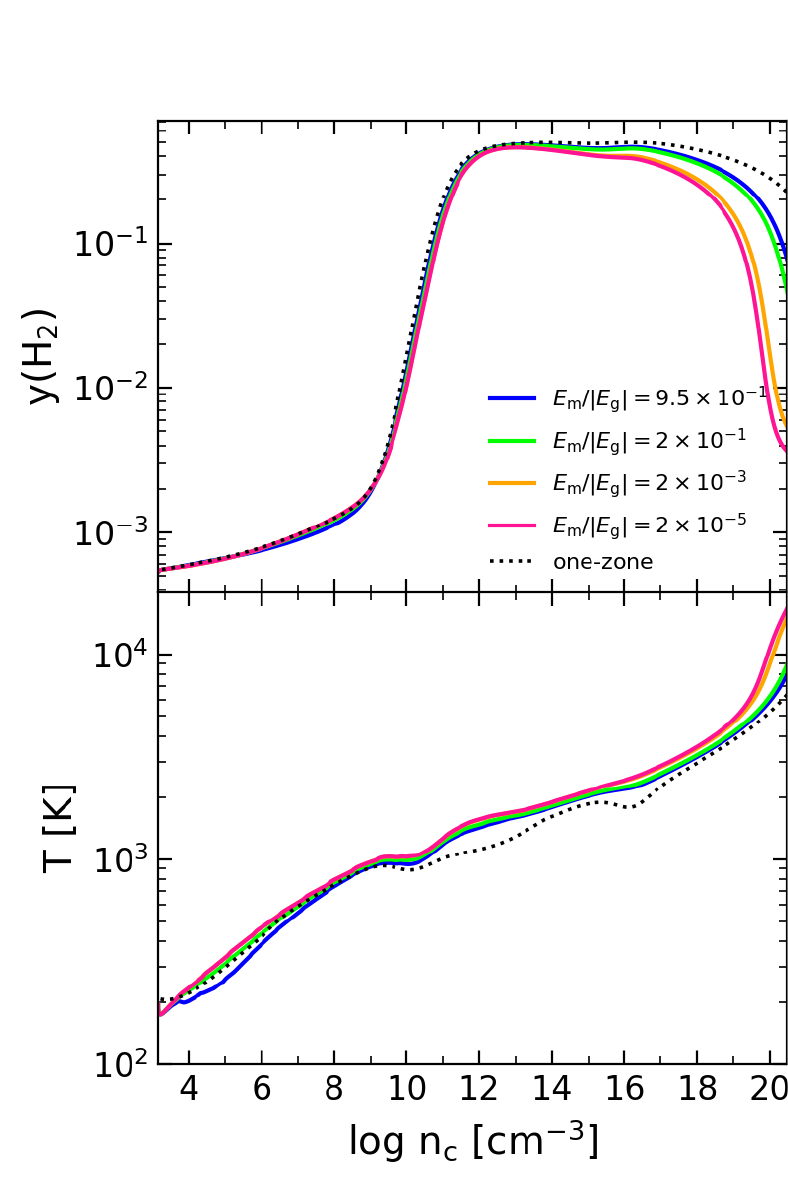}
\caption{
Evolution of the $\rm H_{2}$ abundance (top) and temperature (bottom) of the core as a function of the central number density $n_{\rm c}$.
The non-rotating models with $E_{\rm m}/|E_{\rm g}| = 2\times10^{-5}$ (magenta), $2\times10^{-3}$ (yellow), $2\times10^{-1}$ (green), and $0.95$ (blue) are shown.
For comparison, the black dotted line is the temperature evolution obtained from the one-zone calculation.
}
\label{fig3}
\end{figure}
%---------------------------------
\begin{figure}
\centering
\includegraphics[trim=0 10 0 48, scale = 1.05, clip]{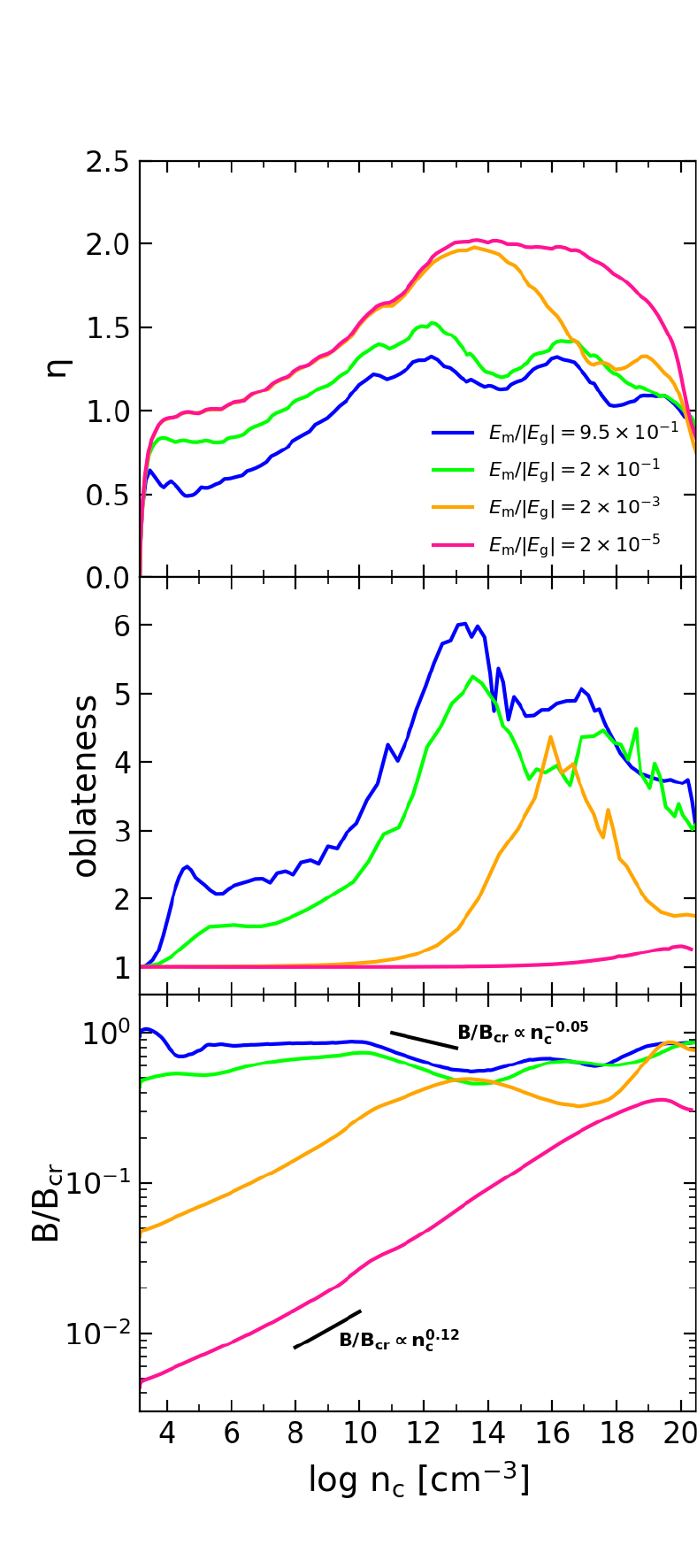}
\caption{
Same as Fig. \ref{fig3},
but for the collapse speed $\eta$ (top), oblateness $\epsilon$ (middle), and magnetic field strength normalized by the critical value $B/B_{\rm cr}$ (bottom).
}
\label{fig4}
\end{figure}
%%%%%%%%%%%%%%%%%%%%%%%%%%%%%%%%%%%%%%%%%%%%%%%%%%%%%%%%%%%%%%%%%%%%%%%
%%%%%%%%%%%%%%%%%%%%%%%%%%%%%%%%%%%%%%%%%%%%%%%%%%%%%%%%%%%%%%%%%%%%%%%
\section{Result}\label{sec3}
\subsection{Evolution of non-rotating magnetized clouds}\label{sec3.1}
First, we study the evolution of non-rotating clouds to see magnetic field effects on the cloud deformation and suppression of the contraction.
We then discuss how much the thermal evolution at the centre is affected by those effects.
Weak and strong field cases are examined in Section \ref{sec3.1.1} and \ref{sec3.1.2}, respectively.
%-----------------------------------
\begin{figure*}
\centering
\includegraphics[trim=0 0 0 0, scale = 0.35, clip]{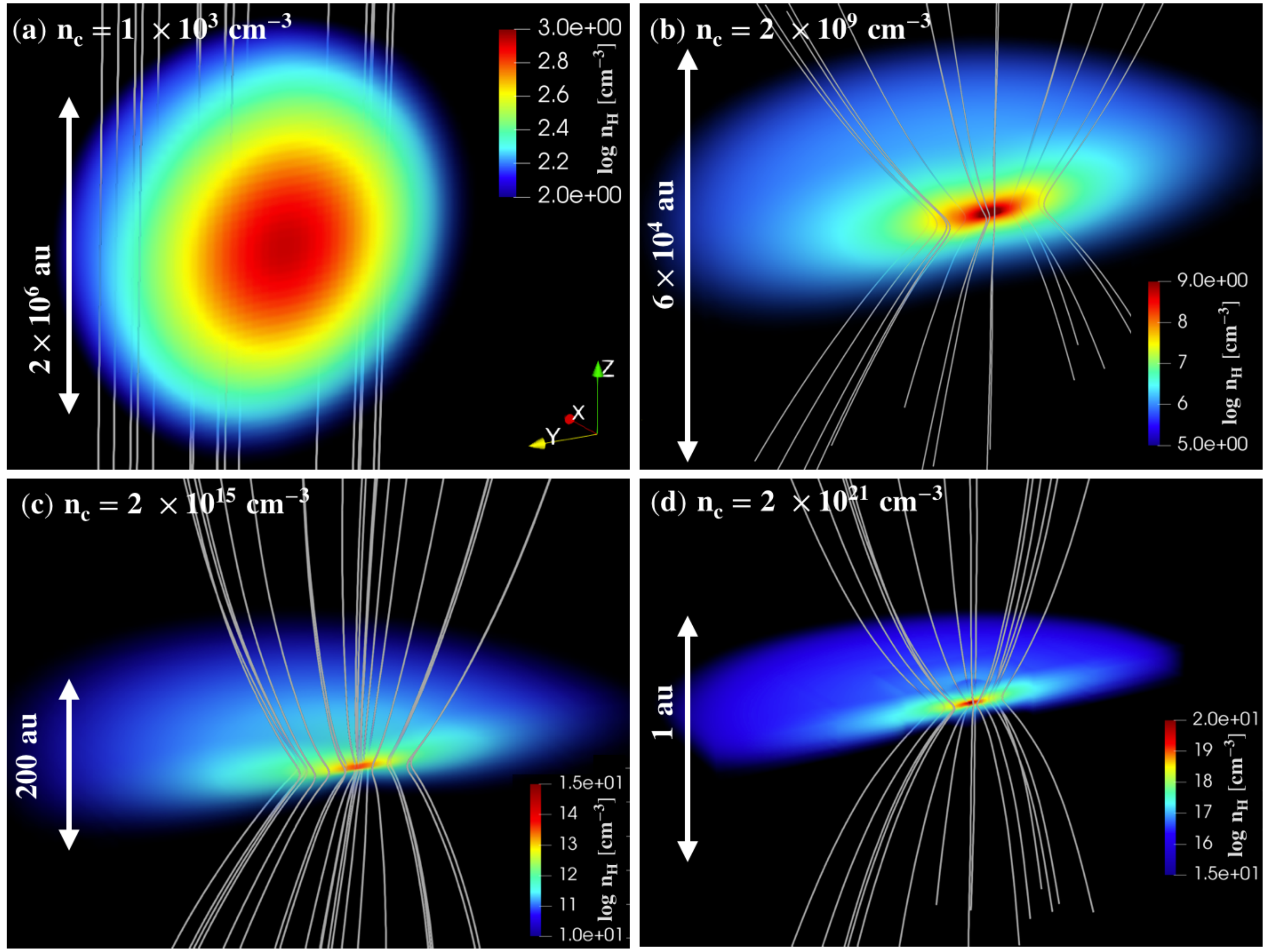}
\caption{
Same as Fig. \ref{fig1}, 
but for the non-rotating strong magnetic field model ($E_{\rm m}/|E_{\rm g}| = 0.95$). 
The central density for panel (d), corresponding to the protostar formation, is $2\times10^{21}\ \rm cm^{-3}$.
}
\label{fig5}
\end{figure*}
%--------------------------------------
\subsubsection{Weak field case}\label{sec3.1.1}
We here see the case with initial magnetic field strength  $E_{\rm{m}}/|E_{\rm g}|=2\times 10^{-3}$, as an example of 
weak field cases. 
 Fig. \ref{fig1} shows the density and magnetic field structure at four stages: (a) the initial state with the central number density $n_{\rm{c}}=1\times10^{3}\ \rm{cm^{-3}}$, (b) $2\times10^{9}\ \rm{cm^{-3}}$, (c) $2\times10^{15}\ \rm{cm^{-3}}$, and (d) $4\times10^{20}\ \rm{cm^{-3}}$, corresponding to the protostar formation.
Fig. \ref{fig2} shows the equatorial (solid) and $z$-axial (dashed) profiles of (a) the density, (b) velocity, (c) temperature, and (d) $\rm{H_{2}}$ abundance 
at the same stages shown in Fig. \ref{fig1}, where physical quantities on the equatorial plane are azimuthally averaged.\par
The initial state is the spherical core with $n_{\rm{c}}=1\times10^{3}\ \rm{cm^{-3}}$ threaded by a weak uniform magnetic field (Fig. \ref{fig1}a).
In early phases, the collapse is proceeding spherically, being unaffected by the weak field (Fig. \ref{fig1}b). 
Since the free-fall timescale $t_{\rm{ff}}\propto n_{\rm c}^{-1/2}$ becomes shorter and shorter with increasing density,
the central collapse proceeds in a runaway fashion with lower density surrounding region left behind.
As a result, the collapsing cloud in this phase consists of a central core with a flat density distribution with the size of a local Jeans length and a lower density envelope with a power-law density distribution with the slope $\propto r^{-2.2}$ (green line in Fig. \ref{fig2}a).
Note that this density slope is slightly steeper than that for the isothermal collapse $\propto r^{-2}$ (\citealp{Larson1969}; \citealp{Penston1969}). 
Since the envelope density becomes $\propto r^{-{2}/{(2-\gamma)}}$ for the polytropic case 
$P\propto \rho^{\gamma}$ (\citealp{Larson1969}; \citealp{Yahil1983}), the density power index of -2.2 corresponds to $\gamma \simeq 1.1$ (\citealp{Omukai_Nishi1998}).\par
Temperature evolution at the centre is shown in Fig. \ref{fig3} (bottom panel) as a function of the density,
along with those in cases with different magnetic field strength.  
The abundance of H$_2$, the dominant coolant in the primordial gas, is shown in the top panel of Fig. \ref{fig3}. 
We can see that the temperature increases gradually with the effective ratio of specific heat 
$\gamma \equiv$ $\rm{d}$ $(\ln{p})/$ $\rm{d}$ $(\ln{\rho})\simeq 1.1\ (T\propto \rho^{0.1})$ before the formation of the adiabatic core, i.e., the protostar, at $\ga 10^{20}\ \rm{cm^{-3}}$, as is expected from the envelope density structure in Fig. \ref{fig2}(a).\par
This thermal behavior is controlled by H$_2$ chemistry that determines the cooling rate. 
The H$_2$ is first produced via the $\rm{H^{-}}$ channel with 
the abundance $\sim 5 \times10^{-4}$ at $n_{\rm{H}}\la 10^{3}\ \rm{cm^{-3}}$.  
The effective ratio of specific heat $\gamma \simeq 1.1$ at $n_{\rm H}=10^{3}-10^{9}\ \rm{cm^{-3}}$ is 
owing to the balance between the compressional heating and the $\rm{H_{2}}$ cooling (bottom panel of Fig. \ref{fig3}). 
At higher densities $n_{\rm{H}} \ga 10^{9}\ \rm{cm^{-3}}$, 
the three-body H$_2$ formation reactions become effective and 
all the hydrogen is converted to the molecules at $n_{\rm{H}}\sim10^{12}\ \rm{cm^{-3}}$
(top panel of Fig. \ref{fig3}).
Although the cooling is enhanced due to the increased H$_2$ abundance, this is almost compensated by 
the chemical heating associated with the H$_2$ formation and the temperature decreases only slightly around $n_{\rm H} \ga 10^{10}\ \rm{cm^{-3}}$ . 
At $n_{\rm{H}} \ga 10^{11}\ \rm{cm^{-3}}$, some $\rm{H_{2}}$ lines become optically thick and the cooling rate gradually declines. At $n_{\rm{H}}\sim10^{13}\ \rm{cm^{-3}}$ the temperature becomes high enough to dissociate some H$_2$ molecules and the dissociation cooling becomes important.
At somewhat higher densities $n_{\rm{H}}\sim 10^{15}\ \rm{cm^{-3}}$, the $\rm{H_{2}}$ collision-induced emission (CIE) also plays some role in the cooling. 
Finally, at $n_{\rm{H}}\sim 10^{16}\ \rm{cm^{-3}}$, the central region becomes optically thick to the $\rm{H_{2}}$ collision-induced absorption 
and the radiative cooling is strongly suppressed thereafter. 
At this time, rapid $\rm{H_{2}}$ dissociation 
begins and associated cooling 
balances with the compressional heating until the completion of the dissociation. 
After that, a protostar forms at the centre, as we will explain in more detail later.\par
Starting with an initial weak field, the field becomes stronger as the field lines are gathered toward the centre by the gas contraction as seen in Fig. \ref{fig1}(b).
The field strength at the centre is plotted in the bottom panel of Fig. \ref{fig4} (yellow line for this model), 
where the field value is normalized by the critical magnetic field $B_{\rm{cr}}$, i.e., 
the field value required for the magnetic force to balance the gravity: if the magnetic field reaches the critical value, the collapse will be stopped by the magnetic force.
Here the critical magnetic field $B_{\rm{cr}}$ is calculated from the balance between the magnetic force $(\propto(\nabla \times \bm{B})\times \bm{B})$ and gravity acting on a uniform density core with the Jeans size, and is given by
 \begin{equation} \label{eq7}
B_{\rm cr} = \sqrt{\frac{ 4 \pi G M_{\rm J}\rho_{\rm{c}}}{R_{\rm J}}},
\end{equation}
where $M_{\rm{J}}$, $R_{\rm{J}}$ and $\rho_{\rm{c}}$ are the Jeans mass, Jeans radius and central mass density, respectively. 
Note that $B_{\rm{cr}}\propto \sqrt{n_{\rm c}T}$ as $R_{\rm{J}}\propto \sqrt{T/n_{\rm c}}$ and $M_{\rm{J}}\propto R_{\rm{J}}^{3} n_{\rm c}$.
In the case of spherical collapse, a magnetic field increases as $B\propto n_{\rm c}^{2/3}$.
To drive this relation, a uniform magnetic field $B$ and uniform density $\rho_{\rm c}$ (or $n_{\rm c}$) are assumed within the cloud radius $R_{\rm c}$.
Then we consider the spherical collapse with mass and magnetic flux conservation ($\rho_{\rm c}R_{\rm c}^3$ = const., $BR_{\rm c}^2$ = const.).
In this case, the magnetic field increases with decreasing cloud radius $R_{\rm c}$ as $B\propto R_{\rm c}^{-2}$ while the density increases as $n_{\rm c} \propto R_{\rm c}^{-3}$. 
Therefore, the relation between $B$ and $n_{\rm c}$ is given as $B\propto n_{\rm c}^{2/3}$.
If we write $B\propto n_{\rm c}^{\alpha}$,
with the shape parameter $\alpha$ ($\alpha=2/3$ for the spherical collapse, as we have seen above), 
the ratio $B$ and $B_{\rm cr}$ is given by $B/B_{\rm{cr}} \propto n_{\rm c}^{\alpha - \gamma/2}$ for $T \propto n_{\rm c}^{\gamma-1}$.
Substituting $\alpha =2/3$ for the spherical collapse and $\gamma = 1.1$ for the primordial gas, 
we obtain the relation $B/B_{\rm{cr}} \propto n_{\rm c}^{0.12}$, shown by the thick black line in the bottom panel of Fig. \ref{fig4}. 
As seen in this figure, the magnetic field grows monotonically along this slope in 
a weak field case (yellow line), which indicates the collapse proceeds spherically.\par
As the field strength approaches the critical value, magnetic effects become apparent in two ways, 
suppression of the contraction and deformation of the cloud.
To see the effect on the contraction, in the top panel of Fig. \ref{fig4}, we show the collapse speed at the centre
with respect to the free-fall time, 
$\eta = t_{\rm{ff}}/t_{\rm{dyn}}$, where $t_{\rm{dyn}}=\rho_{\rm c}/\dot{\rho}_{\rm c}$ is the timescale of the actual central contraction on the simulation, and $t_{\rm{ff}}=\sqrt{3\pi/(32G \rho_{\rm c})}$ is the free-fall time.
For the free-fall collapse with no pressure effect, the collapse speed asymptotically reaches $\eta = 3\pi/2=4.7$,
whereas for the Larson-Penston similarity solution (Larson 1969, Penston 1969), which describes dynamical collapse of an isothermal cloud, it gives $\eta$ = 3.0.
\citet{Omukai2010} have found $\eta$ reaches $\simeq 2.5$ during the prestellar collapse of the primordial gas from one-dimensional hydrodynamical calculation. 
In Fig. \ref{fig4}, the collapse speed for the case with the weakest field ($E_{\rm{m}}/|E_{\rm g}|=2\times 10^{-5}$, magenta line) is unaffected by the magnetic force throughout the collapse. 
The difference in $\eta$ from this curve thus shows the delay due to the magnetic force.
In the case currently discussed ($E_{\rm{m}}/|E_{\rm g}|=2\times 10^{-3}$, yellow), 
the magnetic field does not affect the collapse speed below $n_{\rm{c}} \sim 10^{12}\ \rm{cm^{-3}}$, 
but it begins to delay the collapse once $B/B_{\rm{cr}}$ reaches $\simeq 0.4$ at $n_{\rm{c}}\sim 10^{12}\ \rm{cm^{-3}}$. \par
Simultaneously the cloud shape is deformed by the magnetic force. 
To see the degree of deformation, we plot the oblateness of the core $\epsilon$ in 
the middle panel of Fig. \ref{fig4}. 
The oblateness is defined as $\epsilon = \sqrt{L_{x} L_{y}}/L_{z}$, where $L_{x},\  L_{y},\  L_{z}$ are the $x,\ y,\ z$ axis lengths from the centre to the surface where the density is one-tenth of the central value.
In the model with $E_{\rm m}/|E_{\rm g}|=2\times 10^{-3}$ (yellow), the cloud keeps spherical shape early on ($\epsilon \simeq 1$), and 
begins to become oblate at $n_{\rm{c}}\sim 10^{12}\ \rm{cm^{-3}}$. 
Density and magnetic-field structure of the deformed cloud at $n_{\rm{c}}=2\times 10^{15}\ \rm{cm^{-3}}$ is shown in Fig. \ref{fig1}(c).
As the magnetic force acts mainly on the direction perpendicular to the field lines, 
the contraction is more suppressed in the equatorial rather than in the $z$-axial direction. 
This can be seen in the difference between the equatorial and vertical velocity, i.e., $|v_{r}|<|v_{z}|$ (Fig. \ref{fig2}b, blue line). 
The cloud eventually becomes very oblate, disc-like shape. 
The cloud deformation slows down the field growth 
and $B/B_{\rm cr}$ even decreases with $\propto n_{\rm c}^{-0.05}$ when the centre density reaches $n_{\rm{c}} \simeq 10^{13}\ \rm{cm^{-3}}$(bottom panel of Fig. \ref{fig4}, yellow line) . 
This approximately corresponds to the sheet-like contraction with a scale height $H=c_{\rm{s}}/\sqrt{G \rho_{\rm{c}}}$, where $c_{\rm{s}}$ is the sound speed.
In such a case, we can get $B\propto n_{\rm c}^{1/2}$ from $B\propto R_{\rm c}^{-2}$ (flux conservation) and $n_{\rm c}\propto R_{\rm c}^{-4}$ (mass conservation: $n_{\rm c}R_{\rm c}^2H=$ const.). 
In other words, when the cloud is sheet-like, suppression of the collapse by the magnetic field becomes weaker as the collapse proceeds. 
As a result, magnetic force does not completely stop the contraction.\par
The cloud structure at protostar formation is shown in Fig. \ref{fig1}(d).
We can see a disc-like structure around the protostar. This is a pseudo-disc, 
which is supported by anisotropic magnetic tension, rather than by rotation (\citealp{Galli_Shu1993}).
Around the surface of the pseudo-disc,
the motion in the $z$ direction suddenly decelerated at the shocks located around $z \simeq 0.1$ and $0.6\ \rm{au}$ (Fig. \ref{fig2}b, red line).
As a result, these shocks heat up the gas (Fig. \ref{fig2}c, red line).
Since the shock suppresses the contraction in the $z$ direction, $v_{r}$ becomes larger than $v_{z}$ at $z \simeq 0.1\ \rm{au}$ (Fig. \ref{fig2}b).
The contraction proceeds only in the equatorial direction with the height in the $z$ direction kept constant.  
In such a case, we obtain $B\propto n_{\rm c}$ from $B\propto R_{\rm c}^{-2}$ (flux conservation) and $n_{\rm c}\propto R_{\rm c}^{-2}$ (mass conservation).
This allows the magnetic field to grow again towards the critical value $B_{\rm{cr}}$ at $n_{\rm{c}}\sim 10^{18}\ \rm{cm^{-3}}$ (yellow line in bottom panel of Fig. \ref{fig4}).\par
When the density reaches $n_{\rm{c}}\sim 10^{19}\ \rm{cm^{-3}}$ at the centre, molecular hydrogen is rapidly dissociated due to high temperature, with its abundance falling below $10^{-2}$ (yellow line in top panel of Fig. \ref{fig3}).
The dissociation being almost completed, the temperature increases adiabatically, i.e., $\gamma > 4/3$ thereafter (yellow line in bottom panel of Fig. \ref{fig3}).
As a result, increasing thermal pressure overcomes the gravity and a hydrostatic core, or i.e., a protostar, is formed at the centre.
The protostar has mass $M_{\rm{p}}=2.6\times10^{-3}\ M_{\odot}$ and radius in the $x$-$y$ plane $r_{\rm{p}}=1.4\times10^{-2}\ \rm{au}$.
In our analysis, we have defined a protostar as a core whose central density is greater than $10^{20}\ \rm{cm^{-3}}$ and $\rm{H_{2}}$ abundance is less than $10^{-2}$.
%----------------------------
\begin{figure}
\centering
\includegraphics[trim=0 35 0 65, scale = 1.00, clip]{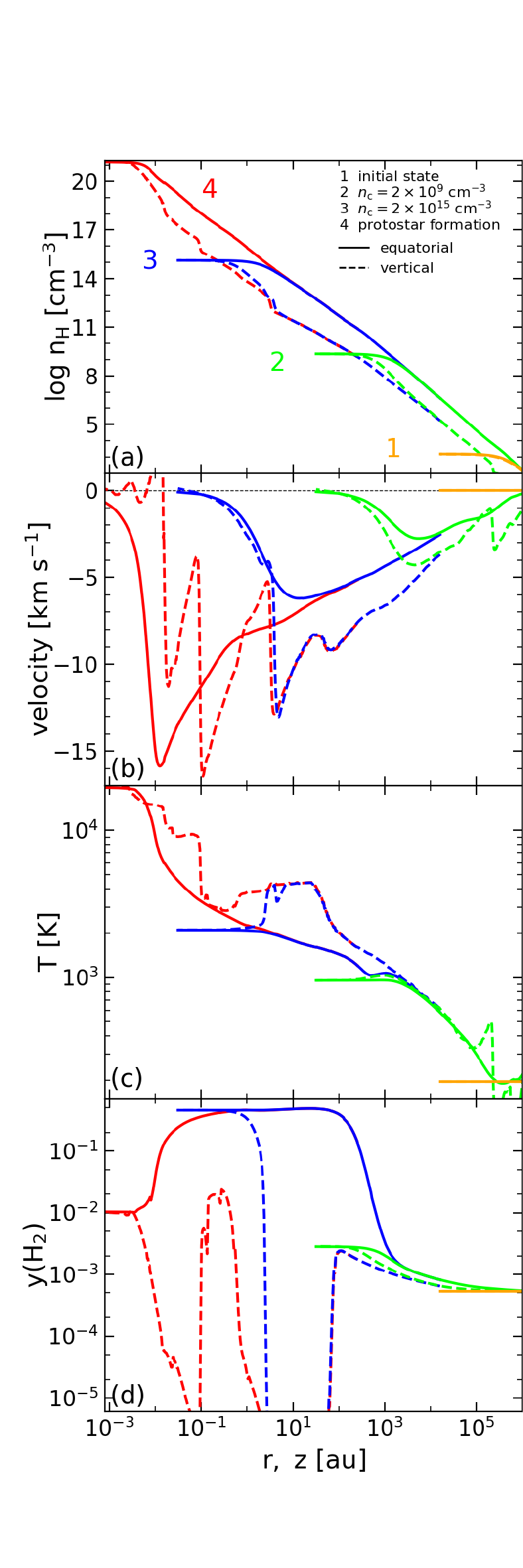}
\caption{
Same as Fig. \ref{fig2}, but for the non-rotating strong magnetic field model ($E_{\rm m}/|E_{\rm g}| = 0.95$). 
The colors correspond to the four epochs in Fig. \ref{fig5}.
}
\label{fig6}
\end{figure}
%-------------------------------------
\begin{figure}
\centering
\includegraphics[trim =0 0 0 0, scale = 1.11, clip]{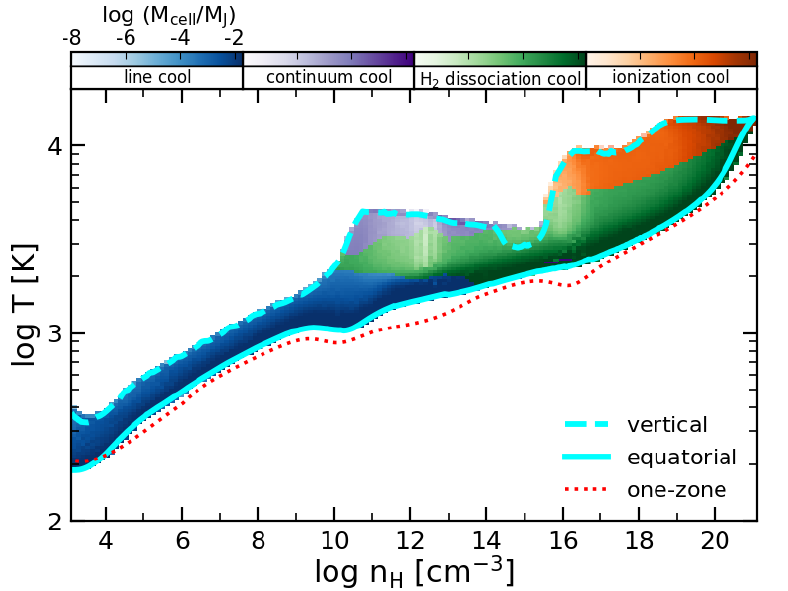}
\caption{
Density-temperature distribution at the epoch of protostar formation in the $n$-$T$ plane for the strong magnetic field model ($E_{\rm m}/|E_{\rm g}| = 0.95$).
The blue, purple, green and orange colors represent the region dominated by line cooling, continuum cooling, $\rm H_{2}$ dissociation cooling and ionization cooling, respectively.
Each color scale indicates the concentration of mass, the cell mass $M_{\rm cell}$ normalized by Jeans mass $M_{\rm J}$, in each bin.
The solid and dashed lines represent the temperature profile in the equatorial and vertical directions.
For comparison, the red dotted line is the temperature obtained from the one-zone calculation.
}
\label{fignT}
\end{figure}
%--------------------------------
\subsubsection{Strong field case} \label{sec3.1.2}
Next, we see the case of a cloud with an initial strong magnetic field comparable to the gravity, i.e.,  $E_{\rm{m}}/|E_{\rm g}|=9.5\times10^{-1}$.
Fig. \ref{fig5} shows the three-dimensional structure of density and magnetic field lines at four evolutionary stages, as in Fig. \ref{fig1}. 
Fig. \ref{fig6} shows the profiles of (a) density, (b) velocity, (c) temperature, and (d) $\rm{H_{2}}$ abundance for the same four stages. \par
Since the initial field is already close to the critical value $B_{\rm{cr}}$ (blue line in the bottom panel of Fig. \ref{fig4}), magnetic field effects are visible soon after the onset of collapse. 
The collapse remains slow with $\eta=0.5-0.6$ until $n_{\rm{c}}\sim 10^{8}\ \rm{cm^{-3}}$ 
(blue line of the top panel of Fig. \ref{fig4}).
This slower collapse and thus lower compressional heating result in lower temperature 
than in the cases with weaker field (Fig. \ref{fig3}, bottom) although the temperature difference is only modest due to high temperature sensitivity of the $\rm{H_{2}}$ cooling rate.\par
Also the cloud shape immediately becomes oblate with $\epsilon \simeq 2$ by the strong magnetic force (blue line in the middle panel of Fig. \ref{fig4})
and thus the disc-like structure forms earlier than weaker field cases. 
Such shape evolution can be also seen in Fig. \ref{fig5}.
As a result of the deformation, the field strength remains at $B/B_{\rm{cr}}\simeq 0.6-0.8$ despite the contraction,
as discussed in Section \ref{sec3.1.1} (blue line in the bottom panel of Fig. \ref{fig4}).\par
When the density reaches $n_{\rm{c}}\sim 10^{15}\ \rm{cm^{-3}}$ (Fig. \ref{fig5}c), 
the collapse speed is still 1.7 times slower than in the weakest field cases ($E_{\rm m}/|E_{\rm g}|=2\times10^{-5}$) 
due to the magnetic force (top panel of Fig. \ref{fig4}).
As a result, the strong field reduces the compressional heating without changing the $\rm H_{2}$ CIE cooling rate 
and the temperature becomes slightly lower compared with the weaker field cases (bottom panel of Fig. \ref{fig3}).
This results in later onset of the H$_2$ dissociation and slightly higher $\rm{H_{2}}$ abundance (upper panel of Fig. \ref{fig3}).  
Consequently the strong magnetic field raises the density at which the protostar forms due to the delayed $\rm H_{2}$ dissociation (Fig. \ref{fig2}a, Fig. \ref{fig6}a).
The stage of the protostar formation is shown in Fig. \ref{fig5}(d),
and we can see the protostellar surface at $r,\ z=0.01\ \rm au$ from the velocity profile (red line in Fig. \ref{fig6}b).\par
Fig. \ref{fignT} shows the density-temperature distribution at the same stage as in Fig. \ref{fig5}(d) in the $n$-$T$ plane.
The blue, purple, green and orange colors represent the region dominated by line cooling, continuum cooling, $\rm H_{2}$ dissociation cooling and ionization cooling, respectively.
The solid and dashed lines indicate the equatorial ($x$-$y$ plane) and vertical ($z-$axis) temperature profiles.
From this figure, we can see that the temperature in regions above the mid-plane is higher than the one-zone calculation (red dotted line).
This is caused by the high compressional heating rate and the shock heating around the pseudo-disc.
The first ($n_{\rm H}\sim 10^{10}\ \rm cm^{-3}$) and second ($n_{\rm H}\sim 10^{16}\ \rm cm^{-3}$) bumps in Fig. \ref{fignT} correspond to the shocks at $z\simeq 63$ and $0.1\ \rm{au}$, respectively (Fig. \ref{fig6}b).
The temperature at the lower density shock is limited by the continuum cooling, especially, the $\rm H^{-}$ free-bound cooling, 
while that at higher density is limited by the ionization cooling.
This result indicates that the temperature in the outer part can be significantly changed by the cloud deformation due to magnetic forces.\par
In summary, magnetic force delays the collapse, as well as changes the cloud shape.
The latter effect prevents the field from reaching the critical value $B_{\rm{cr}}$ and stopping the collapse.
As a result, the change in the temperature evolution due to the delayed collapse remains small at the cloud centre 
even in cases with a strong field. 
On the other hand, pseudo-disc formation by the magnetic force causes shocks near the disc surface, and associated heating significantly modifies the surrounding temperature structure (Fig. \ref{fig6}c or Fig. \ref{fignT}).
%---------------------
\begin{figure}
\centering
\includegraphics[trim=0 20 0 45, scale = 1.05, clip]{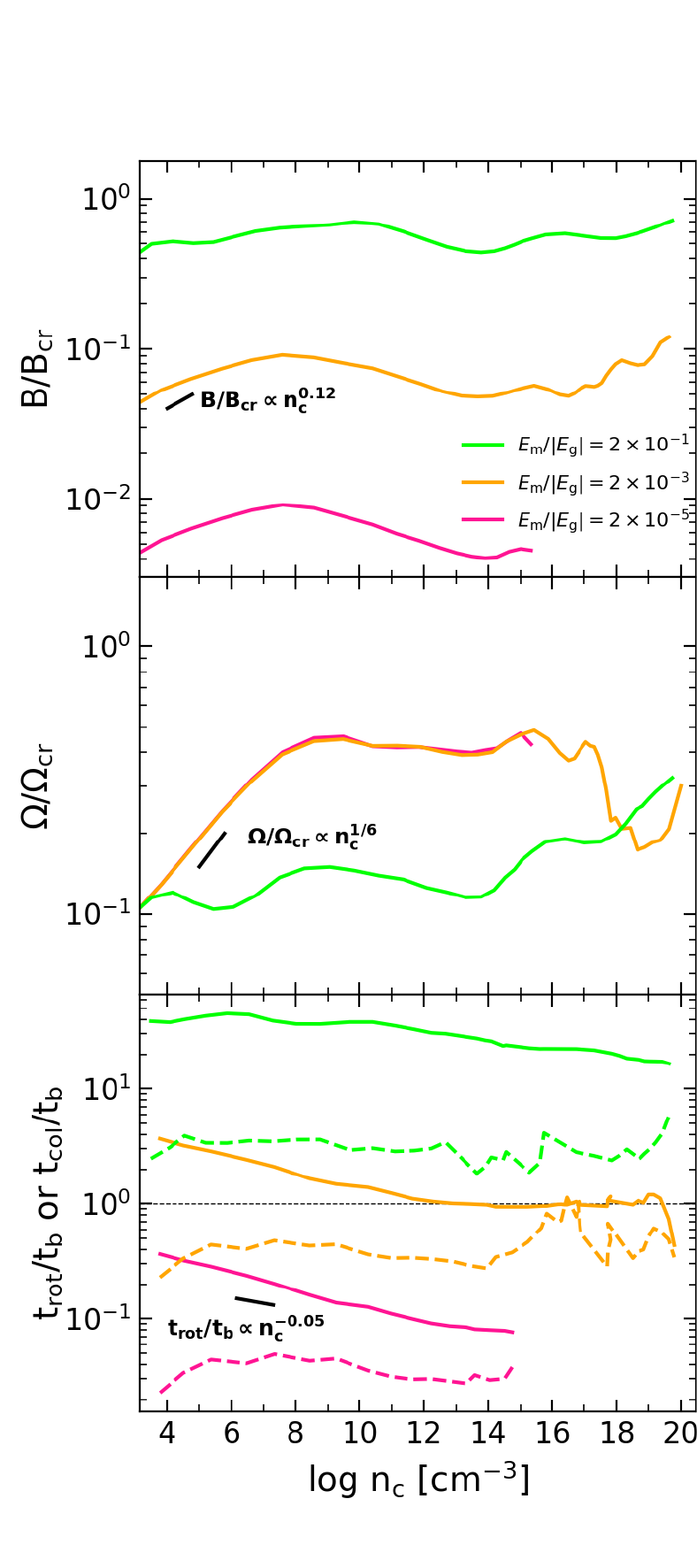}
\caption{
Evolution of the magnetic field normalized by the critical value $B/B_{\rm cr}$ (top) and 
angular velocity normalized by the Keplerian value $\Omega/\Omega_{\rm cr}$ (middle) of the core as a function of the central number density $n_{\rm c}$. 
The rotating models with 
$E_{\rm r}/|E_{\rm g}| = 1\times10^{-2}$ and $E_{\rm m}/|E_{\rm g}| = 2\times10^{-5}$ (magenta), $2\times10^{-3}$ (yellow), and $2\times10^{-1}$ (green) are shown. At the bottom, we also plot the timescale ratios $t_{\rm rot}/t_{\rm b}$ (solid) and $t_{\rm col}/t_{\rm b}$ (dashed),
where $t_{\rm rot}$, $t_{\rm col}$, and $t_{\rm b}$ are the rotational, collapse, and magnetic braking timescale, respectively.
}
\label{fig7}
\end{figure}
%------------------------
\begin{figure*}
\centering
\includegraphics[width=17cm]{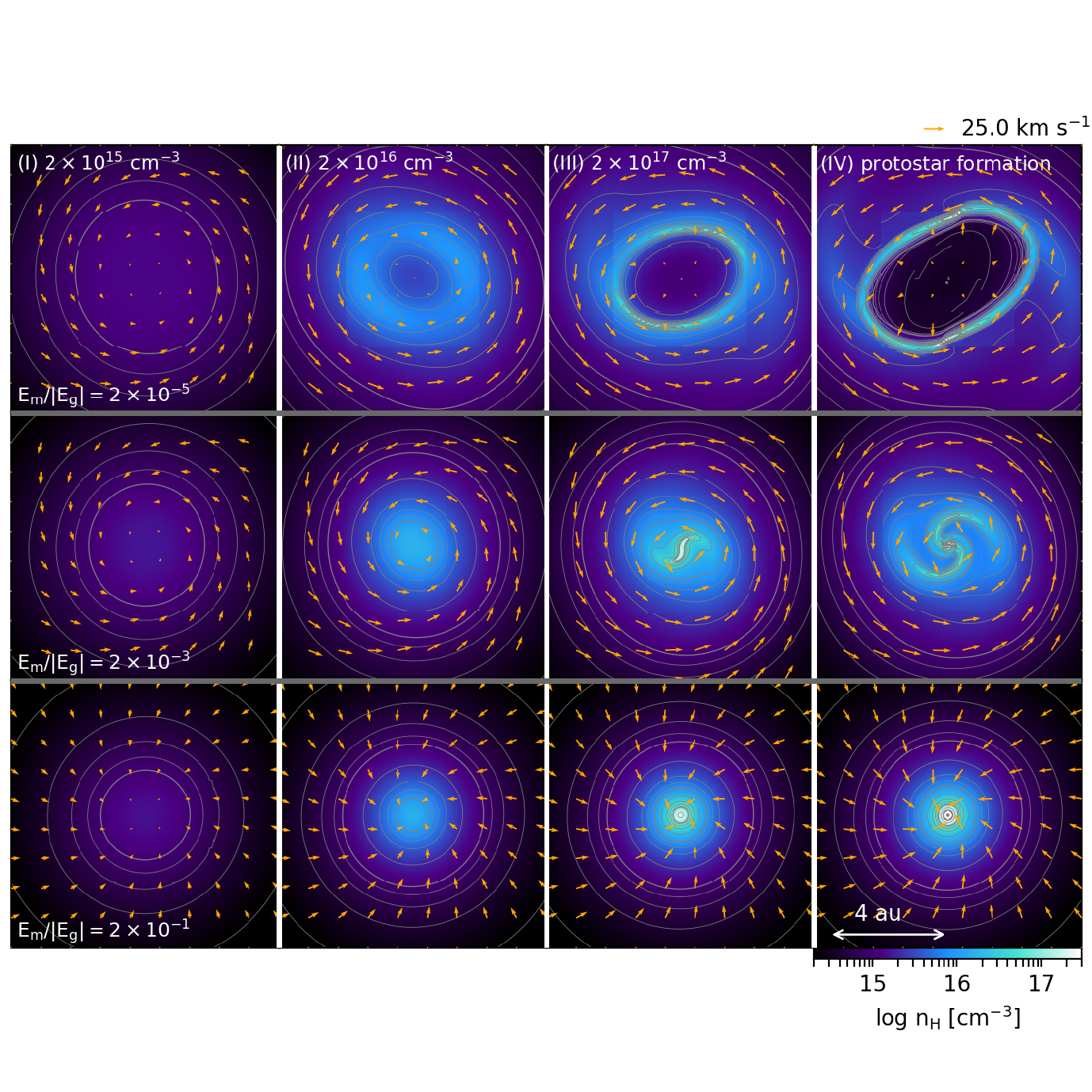}
\caption{
Face-on sliced density distributions at four late stages for the rotating models with $E_{\rm r}/|E_{\rm g}| = 1\times10^{-2}$ and 
$E_{\rm m}/|E_{\rm g}| = 2\times10^{-5}$ (top), $2\times10^{-3}$ (middle), and $2\times10^{-1}$ (bottom).
The first three columns correspond to the stages when the central density 
$n_{\rm c} =$ ($\rm I$) $2\times10^{15}\ \rm cm^{-3}$,  ($\rm I\hspace{-.1em}I$) $2\times10^{16}\  \rm cm^{-3},$ 
and ($\rm I\hspace{-.1em}I\hspace{-.1em}I$) $2\times10^{17}\ \rm cm^{-3}$, with the last column for ($\rm I\hspace{-.1em}V$) just after the protostar formation. Orange arrows in each snapshot represent the projected velocity. The thin lines represent the density contours.
}
\label{fig8}
\end{figure*}
%-------------------------------
\begin{figure*}
\centering
\includegraphics[width=17cm]{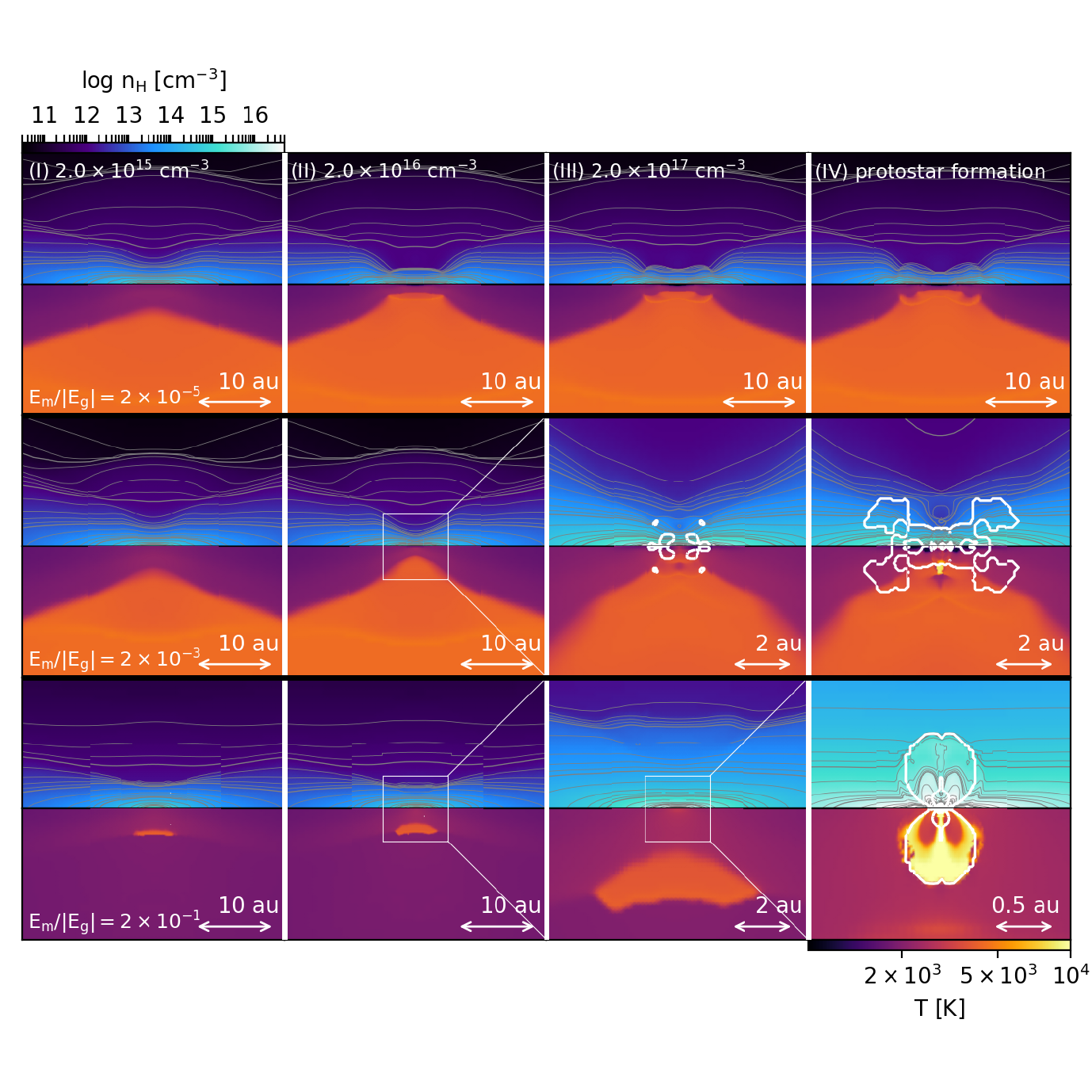}
\caption{
Same as Fig. \ref{fig8}, but for the edge-on sliced density and temperature distributions, which are shown in the upper and lower halves of each panel, respectively.
Thick white contours indicate the outflow regions where the $z$-component of the velocity is outward.
The thin lines in the upper half of each panel represent the density contours.
}
\label{fig9}
\end{figure*}
%-----------------------------------
\begin{figure}
\centering
\includegraphics[width=8cm]{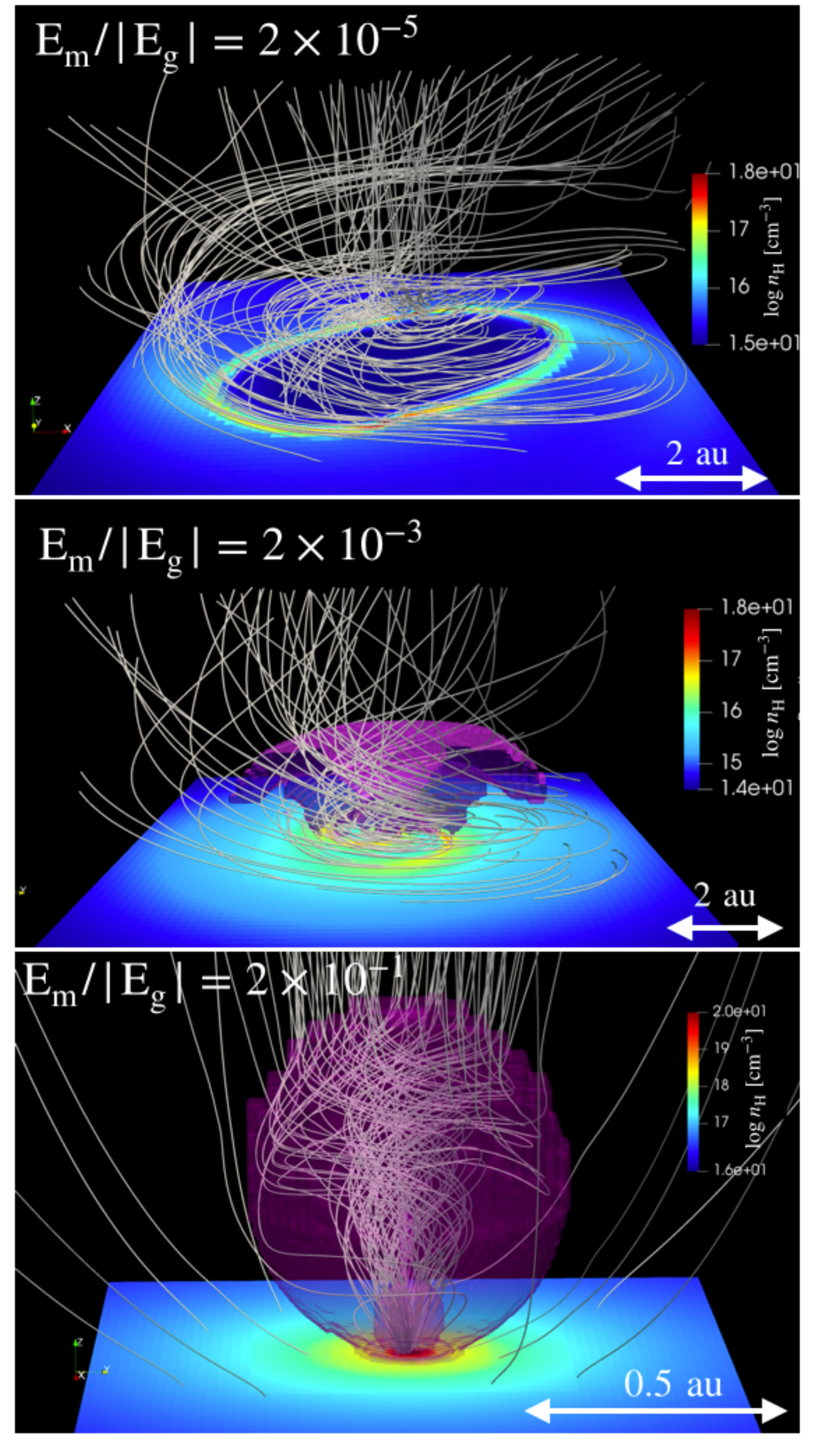}
\caption{
3D structure of magnetic field lines along with the sliced equatorial density distributions just after the protostar formation.
We show the snapshots for the rotating models with $E_{\rm r}/|E_{\rm g}| = 1\times10^{-2}$ and $E_{\rm m}/|E_{\rm g}| = 2\times10^{-5}$ (top), 
$2\times10^{-3}$ (middle), and $2\times10^{-1}$ (bottom).
The purple surfaces indicate the outflow regions.
}
\label{fig10}
\end{figure}
%---------------------------------
%%%%%%%%%%%%%%%%%%%%%%%%%%%%%%%%%%%%%%%%%%%%%%%%%%%%%%%%%%
\subsection{Cases both with rotation and magnetic field}\label{sec3.2}
In this section, we see the cases of rotating and magnetized clouds.
Without magnetic fields, the angular momentum is not transported from the core during axisymmetric collapse. 
Eventually, the centrifugal force balances with the gravity 
and the disc or ring-like structure is formed at the centrifugal radius.  
However, magnetic fields, if exist, can efficiently transport angular momentum. 
For example, in the so-called magnetic braking, 
the field lines twisted by rotation brake the rotational motion by magnetic tension (e.g., \citealp{Mouschovias_Paleologou1979}). 
Also the magnetic field can launch MHD winds from a rotating cloud, thereby also extracting the angular momentum
(\citealp{Blandford_Payne1982}).  
We here investigate the effects of the magnetic braking and MHD outflow on the collapsing cloud.\par
We show the collapse of the cloud with initial rotational energy $E_{\rm{r}}/|E_{\rm g}|=1\times 10^{-2}$ as an example where rotation effects can be seen clearly.  
Note that first-star forming clouds have rotational energy about $E_{\rm{r}}/|E_{\rm g}|=10^{-2}-10^{-1}$ according to the cosmological simulation (e.g., \citealp{Hirano2014}).
Models with different rotational energy will be discussed in Section \ref{sec3.3}.\par
To see the importance of magnetic and centrifugal forces, we show in Fig. \ref{fig7} the magnetic field normalized by the critical value $B/B_{\rm{cr}}$ (top panel) and the angular velocity $\Omega$ normalized by the Keplerian value $\Omega_{\rm{cr}}$ (middle panel) for the models with different initial magnetic field strength $E_{\rm{m}}/|E_{\rm g}|=2\times 10^{-5},\ 2\times 10^{-3},$ and $2\times 10^{-1}$.
We define the Keplerian angular velocity $\Omega_{\rm{cr}}$ as 
%-------------------------------------
\begin{equation} \label{eq8}
\Omega_{\rm{ cr}} = \sqrt{\frac{4\pi G \bar{\rho}}{3}},
\end{equation}
%------------------------------------------
where $\bar{\rho}$ is the average density inside the core, defined as $\rho>\rho_{\rm c}/10$.\par
For later convenience, we describe the increase of angular velocity with density as $\Omega \propto n_{\rm c}^{\alpha}$,
with the index $\alpha$ depending on the cloud shape. 
The value of $\alpha$ can be obtained from the mass and angular momentum conservation: $\alpha=2/3$ ($1/2$) for the spherical (sheet-like, respectively) case.
Note that this dependence is the same as that of the magnetic field described in section \ref{sec3.1}.
Thus, we can get the relation $B/B_{\rm cr} \propto n_{\rm c}^{\alpha-\gamma/2}$ (Section \ref{sec3.1}) and $\Omega/\Omega_{\rm cr} \propto n_{\rm c}^{\alpha - 1/2}$.\par
In the weak field cases of $E_{\rm{m}}/|E_{\rm g}|=2\times 10^{-5}$ and $2\times10^{-3}$, 
early collapse proceeds spherically unaffected either by the magnetic force or rotation.
Therefore, $B/B_{\rm{cr}}$ and  $\Omega/\Omega_{\rm{cr}}$
grow monotonically with the slope corresponding to $\alpha = 2/3$ (top and middle panels in Fig. \ref{fig7}).
By the time the central density reaches $10^{8}\ \rm{cm^{-3}}$, the rotation velocity grows to $\Omega/\Omega_{\rm{cr}} \simeq 0.4 $, and the centrifugal force begins to affect the collapse.
On the other hand, the magnetic field is still weak and has little effect.
In those cases, the cloud becomes oblate solely by the rotation effect at the density $n_{\rm c}\sim 10^{8}\ \rm cm^{-3}$.
After that, $B/B_{\rm cr}$ and $\Omega/\Omega_{\rm cr}$ evolve with the slope corresponding to $\alpha \sim 1/2$ (Fig. \ref{fig7}).
In contrast, for the strong field model ($E_{\rm{m}}/|E_{\rm g}|=2\times 10^{-1}$), the magnetic field remains around $B/B_{\rm{cr}} \lesssim 1$ throughout the collapse.
Due to the strong field, 
the cloud becomes disc-like 
soon after the onset of the collapse.
In addition, the rotation is suppressed by the magnetic braking.
By the combination of the disc-like collapse and magnetic braking, the rotation velocity remains slow, and the centrifugal force is weaker than either gravity or magnetic forces throughout the collapse.\par
The magnetic braking significantly affects the rotation velocity in the strong field case. 
To see when it operates, 
we compare three timescales: the collapse timescale
$t_{\rm{col}}=3\rho_{\rm{c}}/\dot{\rho}_{\rm{c}}$, rotation timescale $t_{\rm{rot}}=2\pi/\Omega$, and angular momentum transfer timescale by the magnetic braking $t_{\rm{b}}=R_{\rm{J}}/v_{\rm{A}}$, 
where $v_{\rm{A}}(=B/\sqrt{4\pi \rho_{\rm{c}}})$ is Alfven velocity at the centre, respectively.
If $t_{\rm{b}}$ becomes shorter than $t_{\rm{col}}$, 
the magnetic braking effectively transports the angular momentum during the collapse.
In the bottom panel of Fig. \ref{fig7}, 
we plot the ratio $t_{\rm{rot}}/t_{\rm{b}}$ (solid) and $t_{\rm{col}}/t_{\rm{b}}$ (dashed) of the average values inside the core for each model.
In the strong field case ($E_{\rm{m}}/|E_{\rm g}|=2\times 10^{-1}$, green line), $t_{\rm{col}}/t_{\rm{b}} > 1 $ from the beginning.
Thus, the angular momentum is extracted from the centre soon after the collapse begins.
In contrast, in the weaker field cases with $E_{\rm{m}}/|E_{\rm g}|=2\times 10^{-3}$ (yellow line) and $2\times10^{-5}$ (magenta line),
$t_{\rm{col}}/t_{\rm{b}}<1$ and the braking timescale is longer than the collapse timescale.
It means the collapse significantly proceeds before the angular momentum is transported by the magnetic braking.
We can also see the effect of the magnetic braking in the behavior of $t_{\rm{rot}}/t_{\rm{b}}$.
Here, $t_{\rm b}$ can be written as $t_{\rm{b}}\sim R_{\rm{J}}/v_{\rm{A}}$ $\sim (c_{\rm{s}}/v_{\rm{A}})t_{\rm{ff}}$ $= (B_{\rm{cr}}/B) t_{\rm{ff}}$, where we use the relation $B_{\rm cr}=(4\pi k_{\rm b}n_{\rm c}T)^{1/2}=(4\pi \rho_{\rm c})^{1/2}c_{\rm s}$.
For conserved magnetic flux,  $B/B_{\rm{cr}}\propto n_{\rm c}^{\alpha-\gamma/2}$ (Section \ref{sec3.1}) and 
$t_{\rm{b}} \propto n_{\rm c}^{(\gamma-1)/2-\alpha}$, while, for conserved angular momentum,  $t_{\rm{rot}}=2\pi/\Omega \propto n_{\rm c}^{-\alpha}$.
Therefore the ratio $t_{\rm{rot}}/t_{\rm{b}}$ is given as $t_{\rm{rot}}/t_{\rm{b}} \propto n_{\rm c}^{(1-\gamma)/2}$, which depends only on the temperature evolution, i.e., the effective ratio of specific heat $\gamma$ and is independent of the cloud shape factor $\alpha$. 
For the primordial gas ($\gamma \simeq 1.1$), this slope is $-0.05$ for conserved angular momentum.
The slope less steeper than $-0.05$ indicates the angular momentum loss due to the magnetic braking.
In the bottom panel of Fig. \ref{fig7}, we can see that the slope in the weak field cases with $E_{\rm{m}}/|E_{\rm g}|=2\times 10^{-5}$ and $2\times10^{-3}$ is indeed $-0.05$, while 
the slope in the strong field case with $E_{\rm{m}}/|E_{\rm g}|=2\times 10^{-1}$ is flatter owing to the magnetic braking.\par
Next we see the evolution in the late collapse phase with $n_{\rm{c}} >10^{15}\ \rm{cm^{-3}}$, where difference in the cloud structure becomes apparent among the models.
Fig. \ref{fig8} shows the cloud structure in this phase for the cases with $E_{\rm{m}}/|E_{\rm g}|=2\times10^{-5}$ (top), $2\times 10^{-3}$ (middle), and $2\times 10^{-1}$ (bottom).
For each model, we show the snapshots of face-on sliced density ($x$-$y$ plane) at 
$n_{\rm{c}}= \ (\rm{I})\  2\times 10^{15} \ \rm{cm^{-3}},\ (\rm {I\hspace{-.1em}I})\ 2\times 10^{16}\ \rm{cm^{-3}},
\ (I\hspace{-.1em}I\hspace{-.1em}I)\ 2\times 10^{17}\ \rm{cm^{-3}}$ , and ($\rm I\hspace{-.1em}V$) just after protostar formation. 
The orange arrows in each panel represent the velocity vectors projected onto the $x$-$y$ plane.
The snapshots on the $x$-$z$ plane at the same epochs are shown in Fig. \ref{fig9}.
The density distribution is shown in the upper half of each panel while the temperature distribution in the lower half.
We enclosed the regions where the velocity in the $z$ direction is outward, i.e., the outflow regions by white lines.
The final snapshot (panel $\rm I\hspace{-.1em}V$) is taken at 20 days after the protostar formation because further calculations require large computational cost .\par
The dependence on the magnetic field strength is apparent in the late collapse phase. 
In stage ($\rm I$), we can see the axisymmetric discs in all three cases.
In the cases with $E_{\rm{m}}/|E_{\rm g}|=2\times 10^{-5}$ and  $2\times10^{-3}$, the disc is rotationally supported, while,  
in the strong field case with $E_{\rm{m}}/|E_{\rm g}|=2\ \times 10^{-1}$, it is magnetically supported.
This difference is reflected in the velocity field: rotating spiral motions are obvious in the 
weak field cases, whereas 
the gases are moving toward the centre with little rotation in 
the strong field case. 
Difference in the magnetic field strength and thus the angular momentum extraction 
via the magnetic braking significantly changes the density structure.
In the case with $E_{\rm{m}}/|E_{\rm g}|=2\times 10^{-5}$ (top panel), where the angular momentum is roughly conserved 
due to ineffective magnetic braking (see $t_{\rm{col}}/t_{\rm{b}},\ t_{\rm{rot}}/t_{\rm{b}} \ll 1$ in bottom panel of Fig. \ref{fig7}), 
a ring is formed at $n_{\rm{c}} \simeq 1\times 10^{15}\ \rm{cm^{-3}}$.
The ring is gravitationally unstable, and immediately breaks into two fragments (Fig. \ref{fig8}, top panel $\rm I\hspace{-.1em}I\hspace{-.1em}I$).
Each fragment collapses until $n_{\rm{c}}\simeq 8\times 10^{20}\ \rm{cm^{-3}}$, with their separation about $5\ \rm{au}$.
In the stronger field case with $E_{\rm{m}}/|E_{\rm g}|=2\times 10^{-3}$,
the magnetic timescale $t_{\rm b}$ satisfies the condition $t_{\rm{rot}}/t_{\rm{b}}\gtrsim 1$
until the rotation motion slows down the collapse at $n_{\rm{c}} \simeq 1\times 10^{15}\ \rm{cm^{-3}}$ (Fig. \ref{fig7} bottom).
This means that the magnetic braking significantly slows down the rotation.
As a result, we observe the formation of spiral arms instead of a ring at the stage ($\rm I\hspace{-.1em}I\hspace{-.1em}I$).
The central part further contracts and a protostar is formed at $n_{\rm{c}}\simeq 2\times 10^{20}\ \rm{cm^{-3}}$ with the radius $4\times10^{-2}\ \rm{au}$.
In the strongest field case with $E_{\rm{m}}/|E_{\rm g}|=2\times 10^{-1}$, rotational motion is strongly suppressed by the magnetic braking and spherical collapse continues (Fig. \ref{fig8}, bottom) 
until the protostar formation at stage ($\rm I\hspace{-.1em}V$) with the radius $4 \times10^{-2}\ \rm{au}$.
In summary, the strong magnetic field efficiently extracts angular momentum, thereby suppressing fragmentation during the collapse.\par
Finally, we investigate the outflow from the central region.
Fig. \ref{fig10} shows the density and magnetic-field structure just after protostar formation (stage $\rm I\hspace{-.1em}V$ in Fig. \ref{fig9}), where  
the outflow region is indicated by purple surface for the cases with 
$E_{\rm{m}}/|E_{\rm g}|=2\times10^{-5},\ 2\times 10^{-3}$, and $2\times10^{-1}$. 
In the cases with $E_{\rm m}/|E_{\rm g}| \geq 2\times10^{-3}$ , 
the outflows are launched due to the field lines twisted by the rotational motion. 
In the strongest field case with $E_{\rm{m}}/|E_{\rm g}|=2\times 10^{-1}$, 
rotational motion begins to dominate during the protostar formation and generates a toroidal magnetic field ($B_{r},\ B_{\phi}$), 
which launches the outflow (bottom panel of Fig. \ref{fig10}) with the maximum velocity $v_{\rm{max}}=40\ \rm{km\ s^{-1}}$,
in the same order of magnitude as the escape velocity. 
In the moderate case with $E_{\rm{m}}/|E_{\rm g}|=2\times 10^{-3}$,
a rotating disc emerges during the collapse and outflows are launched (middle row at stage $\rm I\hspace{-.1em}I\hspace{-.1em}I$ of Fig. \ref{fig9}) with slower velocity ($v_{\rm{max}}=16\ \rm{km\ s^{-1}}$) owing to the shallower gravitational potential of the launching points than in the former case.
No outflow is observed in the weakest field case with $E_{\rm{m}}/|E_{\rm g}|=2\times 10^{-5}$. 
Because of very weak magnetic force, 
the field lines are dragged by the rotational motion, resulting in a toroidal field with a configuration almost parallel to the disc $B_{z},\ B_{r} \ll B_{\phi}$ (top panel of Fig. \ref{fig10}).
Magneto-centrifugal winds cannot be launched by this kind of field configuration since the centrifugal force is almost perpendicular to the field lines.
If we extend the calculation until later time, however, 
the field could be further amplified by rotation of the disc. 
In such a case, another mechanism of MHD winds, called 
magnetic-pressure-driven wind (e.g., \citealp{Blandford_Payne1982}; \citealp{Tomisaka2002}; \citealp{Machida2008a}), might be eventually launched. \par
In our simulations, the amount of angular momentum transferred by the outflow is smaller than that by the magnetic braking 
since the outflows are launched only in a late phase of the protostar formation.
Also in the present-day star formation, 
the angular momentum is mainly removed by magnetic braking.(e.g., \citealp{Marchand2020}).
We thus expect that the magnetic braking plays a more important role in controlling fragmentation of first-star forming clouds.
%------------------------------------------------
\begin{figure*}
\centering
\includegraphics[trim =0 30 3 30, scale = 1.26, clip]{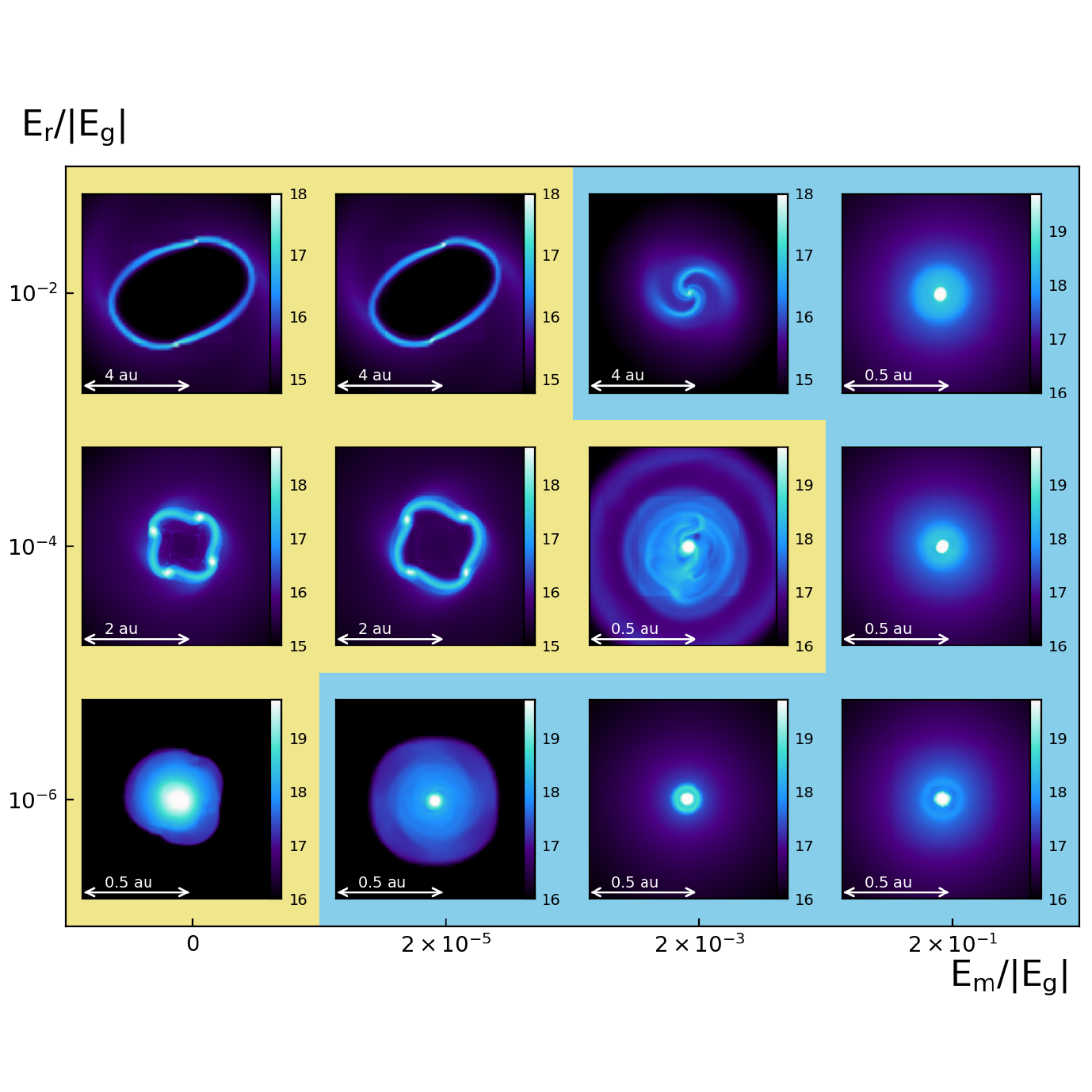}
\caption{
The equatorial density structures just after the protostar formation plotted against $E_{\rm m}/|E_{\rm g}|$ and $E_{\rm r}/|E_{\rm g}|$. 
Frgmentation occurs in the models on the yellow background, whereas the cloud monolithically collapses until the protostar formation in the models on the blue background.
The color scale represents $\log{n_{\rm H}}\ \rm [cm^{-3}]$. 
}
\label{fig11}
\end{figure*}
%-----------------------------------------------------
\begin{figure*}
\centering
\includegraphics[trim=0 35 20 40, scale = 1.32, clip]{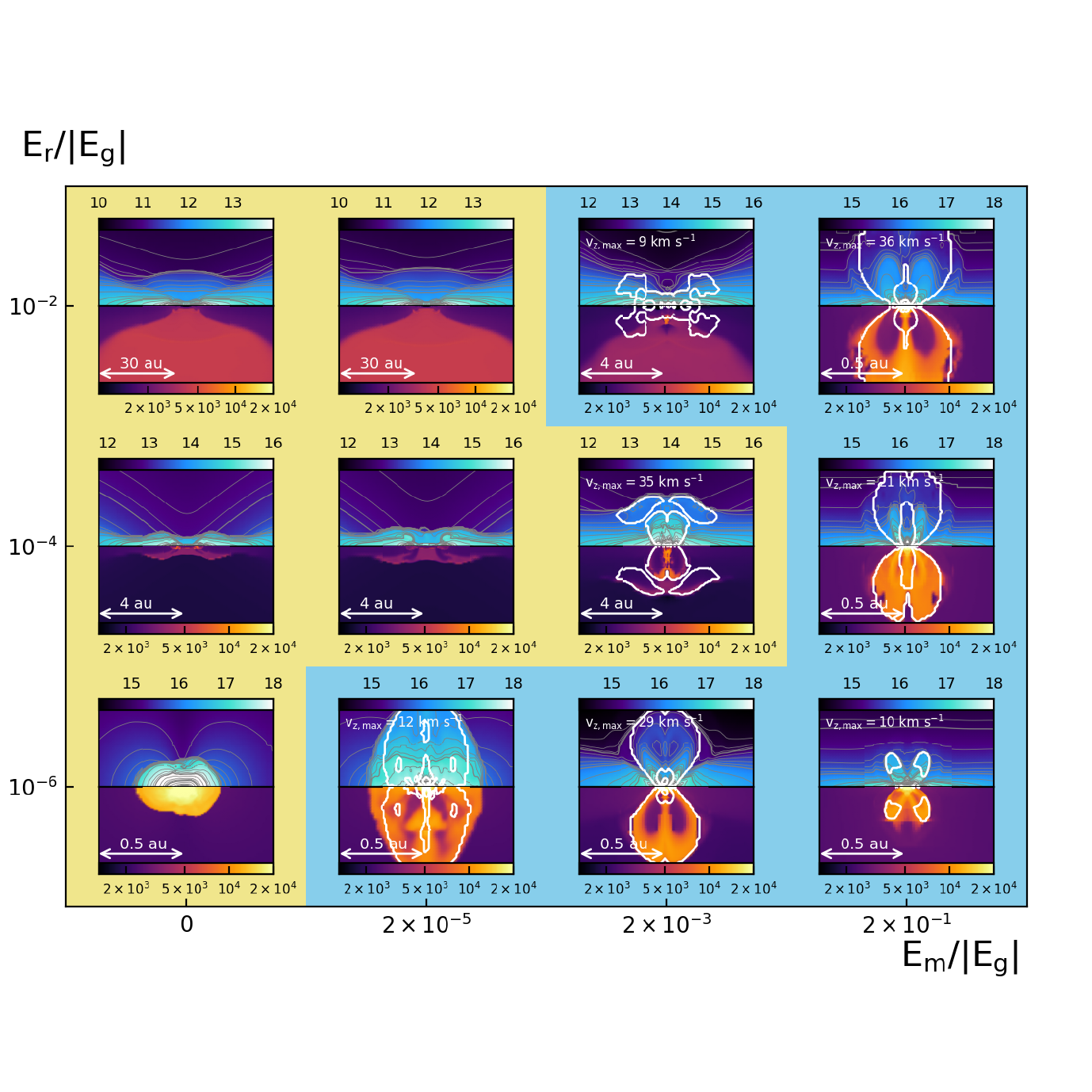}
\caption{
Same as Fig. \ref{fig11}, but for the vertical density and temperature structures. 
The color scales represent $\log{ n_{\rm H}}\ \rm [cm^{-3}]$ (upper) and $T\ \rm [K]$ (lower).
Thick white contours indicate the outflow regions, with the thin lines in the upper half of each panel representing the density contours.
The maximum outflow velocity in the $z$-direction are shown in each panel.
}
\label{fig12}
\end{figure*}
%-----------------------------------------------------------
\begin{figure*}
\centering
\includegraphics[trim =0 30 3 30, scale = 1.26, clip]{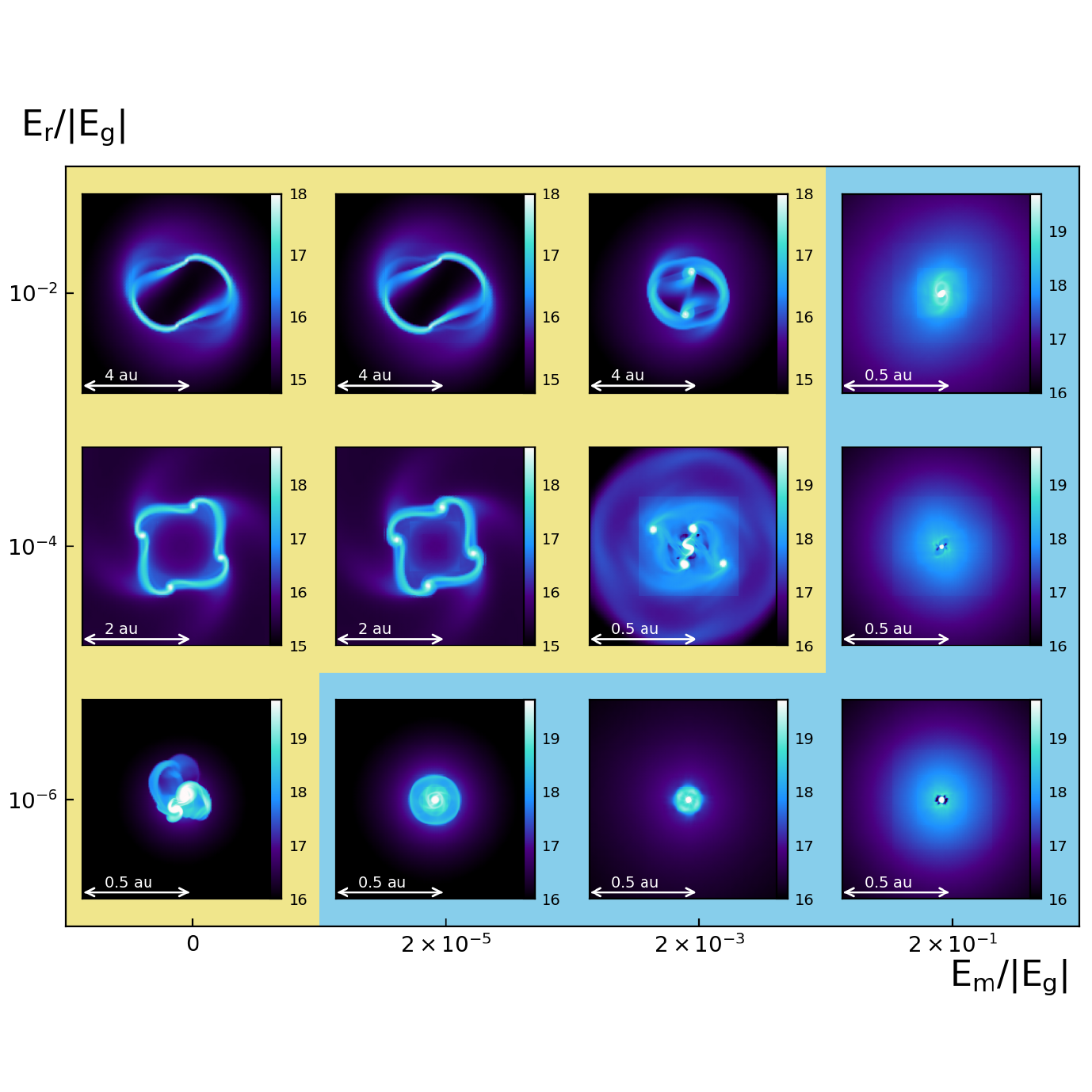}
\caption{
Same as Fig. \ref{fig11}, but for the barotropic simulations.
}
\label{fig13}
\end{figure*}
%-------------------------------------------------------------------
\begin{figure*}
\centering
\includegraphics[trim=0 35 20 40, scale = 0.87, clip]{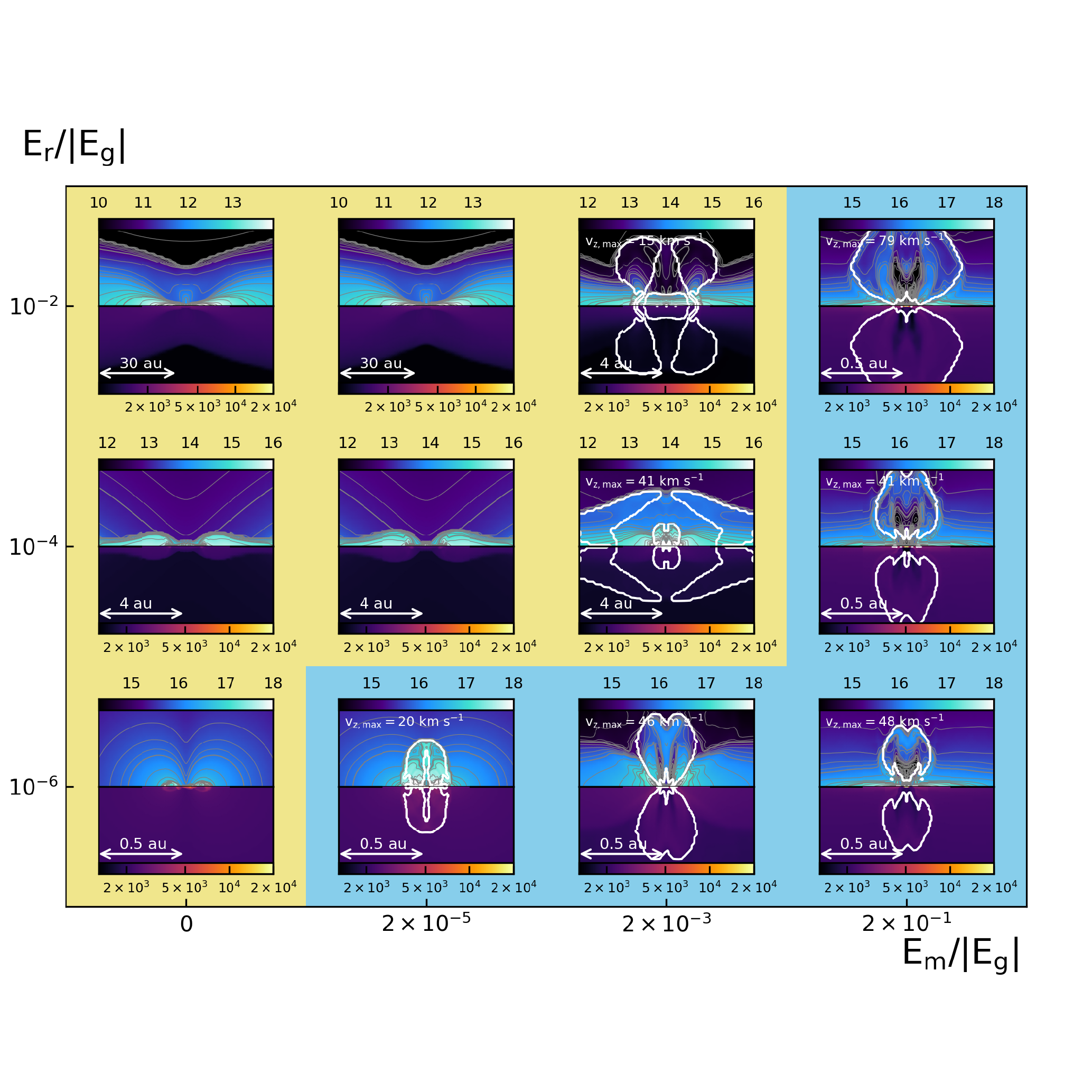}
\caption{
Same as Fig. \ref{fig12}, but for the barotropic simulations.
}
\label{fig14}
\end{figure*}
%------------------------------------------------------------------
\subsection{Fragmentation and outflow conditions}\label{sec3.3}
\subsubsection{Fragmentation conditions}
First, we discuss the fragmentation condition.
The density structure on the $x$-$y$ plane just after the protostar formation is shown in Fig. \ref{fig11} for models with different initial field strength $E_{\rm{m}}/|E_{\rm g}|$ and rotation speed $E_{\rm{r}}/|E_{\rm g}|$. 
The outcome can be classified into two types: the case in which fragmentation is observed during the collapse (yellow background) 
and that in which the collapse proceeds monolithically without fragmentation (blue background).
As seen in Section \ref{sec3.2}, 
rotation facilitates the fragmentation.
Thus, rapidly rotating clouds form a ring-like structure and then fragment.
Also the separation among the fragments tends to be larger for clouds with faster rotation as seen for the cases without magnetic field (leftmost column of Fig. \ref{fig11}). 
In our case, fragmentation takes place through the ring formation due to our choice of small initial density fluctuations of about 1$\%$. 
If large enough density fluctuations are assumed, the fragmentation instead would occur via bar-like structure (\citealp{Machida2008b}).
It is, however, known that the amplitude of initial seed fluctuation does not significantly affect the fragmentation condition while it determines the mode of fragmentation.
A large magnetic field suppresses fragmentation through the angular momentum transport (Section \ref{sec3.2}) and causes monolithic collapse.
As seen in Fig. \ref{fig11}, fragmentation occurs in cases where  
the rotation energy $E_{\rm{r}}$ is larger than the magnetic field energy $E_{\rm{m}}$ and vice versa.
If those two values are close $E_{\rm{r}} \sim E_{\rm{m}}$, fragments, if any, immediately merge due to the small separation.
Here, we denote the models that show such evolution as merger models, 
examples of which are the cases with $(E_{\rm{r}}/|E_{\rm g}|, \ E_{\rm{m}}/|E_{\rm g}|)=(10^{-4},\ 2\times10^{-3})$ and $ (10^{-6},\ 0)$. \par
\subsubsection{Outflow conditions}
Next we see the condition for the outflow launching. 
The density and temperature structure in the $x$-$z$ plane is 
shown in Fig. \ref{fig12} at the same epoch as in Fig. \ref{fig11}.
The thick white lines indicate the outflow regions where the velocity in the $z$ direction is outward.
We can see that outflows are launched in all the monolithic collapse models 
and one merger model $(E_{\rm{r}}/|E_{\rm g}|,\ E_{\rm{m}}/|E_{\rm g}|)=(10^{-4},\ 2\times 10^{-3 })$, where a single protostar is formed (Fig. \ref{fig11}).
Two types of outflow are known: the magneto-centrifugal wind, which is driven by the centrifugal force along magnetic field lines,
and the magnetic-pressure wind,  which is driven by the magnetic-pressure gradient force
(\citealp{Tomisaka2002}; \citealp{Banerjee_Pudritz2006}; \citealp{Machida2008a}).
The magneto-centrifugal wind is launched from a cloud with a strong initial field when the magnetic energy exceeds the thermal energy in the central region (\citealp{Machida2008a}).
In our case, this wind mode becomes dominant for the models with $E_{\rm{m}}/|E_{\rm g}| \geq 2\times10^{-3}$, where the field is sufficiently amplified during the collapse.
In Fig. \ref{fig11}, we can identify V-shaped density structures in these models, which are characteristic for the magneto-centrifugal winds (\citealp{Tomisaka2002}). 
On the other hand, the magnetic-pressure-driven wind can be launched even from clouds with initially weak fields by winding up the field lines many times by the rotation.  
Once enough magnetic pressure gradient force is built up to overcome the gravity, winds can be driven. 
An example can be observed in the case with $(E_{\rm{r}}/|E_{\rm g}|,\ E_{\rm{m}}/|E_{\rm g}|)=(10^{-6},\ 2\times 10^{-5})$, whose structure is similar to the so-called magnetic bubble (\citealp{Tomisaka2002}).
In this case, a large part of the outflow region is occupied by a toroidal rather than a poloidal field,
which is a typical magnetic-field configuration for the magnetic-pressure-driven wind (\citealp{Machida2008a}; \citealp{Tomida2013}).\par
In any magnetic outflow mechanism, the rotational motion is crucial for the outflow launching.
When the gas at a small radius around the protostar has sufficiently large angular momentum, 
the field lines can be immediately wound up and the outflow is launched.
On the other hand, a disc takes time to wind up the field lines.
For this reason, we cannot see the outflow from the disc in the fragmentation model just after the protostar formation.
Also in these models, outflows are not launched from the protostars until the end of our simulation since most of the angular momentum is in the orbital motion.\par
In summary, 
a magneto-centrifugal wind is launched just after the protostar formation if $E_{\rm{m}}/|E_{\rm g}| \geq 2\times10^{-3}$.
Even in the case with a weaker field,
if a single protostar is formed due to the effect of magnetic braking,
rotation of the protostar immediately winds up the field lines, and as a result magnetic-pressure-driven winds can be launched.
Note, however, that even in cases where we did not observe outflows, the magnetic-pressure-driven winds may be driven 
in a later phase if we follow longer term evolution. 
\subsection{Comprison with barotropic simulation}\label{sec3.4}
Finally, we compare the results of our simulations in which the gas dynamics and thermal evolution are consistently solved with those assuming the barotropic equation of state derived from the one-zone calculation, which is frequently adopted in previous works (e.g., \citealp{Machida2008}).
The temperature evolution obtained from the one-zone model is that of the cloud centre calculated under the assumption of nearly free-fall collapse.
In this section, we first examine how much the temperature in the self-consistent simulation deviates from that derived from the one-zone calculation.
We then compare the fragmentation and outflow conditions in the two types of simulations and discuss the validity of the barotropic approximation.
In Fig. \ref{fig13} and Fig. \ref{fig14}, we show the face-on and edge-on snapshots at the protostar formation for the barotropic simulations, as in Fig. \ref{fig11} and Fig. \ref{fig12}, for comparison.\par
First of all, we consider the difference in the temperature structure between these two simulations.
Although the temperature evolution at the centre does not largely deviate from the barotropic case (Fig. \ref{fig3}), 
the temperature in outer regions can be very different due to the effects such as shock heating by the accretion and outflows (Fig. \ref{fig12}).
For example, in the cases of $(E_{\rm{r}}/|E_{\rm g}|,\ E_{\rm{m}}/|E_{\rm g}|)=(10^{-4}, \ 0 )$ and $(10^{-4},\ 2\times 10^{-5})$, the disc surface is heated by the accretion shock.
In the cases of $(E_{\rm{r}}/|E_{\rm g}|,\ E_{\rm{m}}/|E_{\rm g}|)=(10^{-2}, \ 0 ), \ (10^{-2},\ 2\times 10^{-5}),$ and $ (10^{-2},\ 2\times 10^{-3})$, 
hot regions of $4\times10^{3}\ \rm{K}$ spread 
above the disc since the $\rm{H_{2}}$ is destroyed by a shock at $z=70\ \rm{au}$. 
In all the cases with outflows in Fig. \ref{fig12}, the outflow cavities are heated up to $10^4\ \rm K$ by the shocks.
In the model $(E_{\rm{r}}/|E_{\rm g}|,\ E_{\rm{m}}/|E_{\rm g}|)=(10^{-6},\ 0 )$, in which an optically thick ring is formed,
the ring is heated by the shock and deforms to a hot bubble structure.
Needless to say, the barotropic simulations cannot reproduce the shock heating as seen in Fig. \ref{fig12}. \par
Next, we discuss the effect of the temperature difference on the fragmentation.
From the comparison of Fig. \ref{fig11} and Fig. \ref{fig13},
we can see that cases in which fragmentation occur or not are the same,
except for a single case $(E_{\rm{r}}/|E_{\rm g}|,\ E_{\rm{m}}/|E_{\rm g}|)=(10^{-2}, \ 2\times10^{-3} )$.
In this model, the temperature of the disc is cooled by the $\rm{H_{2}}$ CIE and adiabatic expansion and the disc becomes gravitationally unstable.
As a result, the $m=2$-mode fluctuation grows rapidly and a arm structure appears.
Despite a largely different thermal structure in outer regions, the fragmentation condition remains almost the same among the two sets of simulations since the fragmentation tends to happen in the densest core region where the temperature difference is small.  
Note also that the final outcomes of the merger models $(E_{\rm{r}}/|E_{\rm g}|,\ E_{\rm{m}}/|E_{\rm g}|)=(10^{-4}, \ 2\times10^{-3} )$ and $(10^{-6}, \ 0 ) $ differ:
the protostars in the self-consistent simulation have
larger radii owing to the shock heating and merge immediately in the realistic simulation (Fig. \ref{fig11}), 
while they do not in the barotropic simulation (Fig. \ref{fig13}).\par
Finally, we discuss the effect on outflows.
Comparing Fig. \ref{fig12} and Fig. \ref{fig14}, we see again that the cases in which the outflows are present are the same among two sets of simulations, 
even though the surrounding thermal structure is significantly different because of the shock heating.
This implies that the thermal structure near the centre controls the outflow driving.
The strength of the outflow, however, is different: the flows are weaker in the self-consistent simulations.  
For the case with $(E_{\rm{r}}/|E_{\rm g}|,\ E_{\rm{m}}/|E_{\rm g}|)=(10^{-2}, \ 2\times10^{-1} )$, the maximum velocity in the $z$-direction is 
$v_{z,\rm{max}}=36\ \rm{km\ s^{-1}}$ in the self-consistent simulation, while $v_{z,\rm{max}}=79 \ \rm{km\ s^{-1}}$ in the barotropic simulation.
This is because the protostellar radius is larger in 
the self-consistent simulation due to the shock heating 
and thus the depth of the gravitational potential well is shallower.
This results in the smaller escape (and outflow) velocity.
\label{Sec:result}
%%%%%%%%%%%%%%%%%%%%%%%%%%%%%%%%%%%%%%%%%%%%%%%%%%%%%%%%%%%%%%%%%%%%%%%%%%%%%
%%%%%%%%%%%%%%%%%%%%%%%%%%%%%%%%%%%%%%%%%%%%%%%%%%%%%%%%%%%%%%%%%%%%%%%%%%%%
\section{Summary and Discussion}\label{sec4}
We have studied the collapse of magnetized primordial clouds starting from the stage of the dense cores with the central density $n_{\rm{c}}\simeq10^{3}\ \rm{cm^{-3}}$ until the protostar formation ($n_{\rm{c}}\simeq10^{20}\ \rm{cm^{-3}}$) with 3D magnetohydrodynamics simulation incorporating self-consistently the thermal processes and non-equilibrium chemical reactions.
In this work, we have investigated the effect of magnetic fields on the thermal evolution and fragmentation processes.
Besides, we have examined how the fragmentation and outflow launching conditions are altered from simulations using the barotropic approximation. 
Our findings can be summarized as follows: 
\begin{itemize}
\item
Magnetic fields slow down the cloud contraction only in the directions perpendicular to the field lines and not in the parallel directions to them. 
Also even if the field is amplified and becomes close to the critical value $B_{\rm cr}$ for stopping the gravitational contraction, the cloud deformation due to the magnetic force prevents the fields from reaching the critical value. 
The collapse always proceeds nearly at the free-fall rate of the central core and the temperature evolution is hardly altered by the magnetic force even in the strong field cases.  
\item
The magnetic braking is the main mechanism that transports angular momentum from collapsing primordial star-forming clouds. 
If the initial magnetic energy $E_{\rm m}$ is stronger than rotational energy $E_{\rm r}$,
the magnetic braking can transport enough angular momentum to prevent a disc formation and its fragmentation. 
On the other hand, angular momentum transport by the outflows does not affect fragmentation during the pre-stellar collapse since they are launched only in the later phase. 
\item
The fragmentation and outflow launching conditions obtained from our self-consistent simulations are similar to those from the simulations adopting the barotropic equation of state from the one-zone collapse calculation. 
This is because either magnetic or centrifugal force
does not considerably change the central temperature evolution during the collapse, which controls the fragmentation and outflow launching. 
However, when the initial magnetic-field and rotational energies are close, 
differences in the fragmentation processes are observed due to the differences in thermal evolution caused by the shock heating or cooling processes.
\end{itemize}
Some effects are not present in the barotropic calculation as we have already mentioned above. 
For example, the protostellar radius becomes larger due to the shock heating at the stellar surface, which facilitates the merger of protostars. 
Also, after the protostar formation, temperature in the disc may be different from that in the barotoropic simulation.
Although the barotropic approximation is good enough during the collapse phase, 
the gas dynamics and thermal evolution need to be solved consistently for the later phase, 
in which the radiation from the protostar plays an important role as a feedback. \par
First-star forming clouds have rotational energy of about $E_{\rm{r}}/|E_{\rm g}|=10^{-2}-10^{-1}$ acccording to previous cosmological simulations
(e.g., \citealp{Hirano2014}), roughly corresponding to the cases of $E_{\rm{r}}/|E_{\rm g}|=10^{-2}$ in our calculation. 
With this level of rotation, our results indicate that the magnetic field effect becomes important
for $B>10^{-6} ( n_{\rm c}/10^{3}\ {\rm cm^{-3}} )^{2/3}\ \rm G$ 
during the collapse phase, considering the amplification law in the weak field case $B\propto n_{\rm c}^{2/3}$.
Theoretically, the cosmological fluctuations during the epoch of recombination (\citealp{Ichiki2006}) or curved shocks in a minihalo (\citealp{Xu2008})
can produce a uniform primordial magnetic field of at most $10^{-18}\ \rm{G}$ (\citealp{Ichiki2006}).
From these results, we can see that such a uniform week field do not affect the cloud evolution during the collapse phase. \par
If the gas in a minihalo is highly turbulent, however, 
the primordial field can be amplified to the critical value by a small-scale dynamo during the collapse phase (\citealp{Schober2012}) 
and may affect first-star formation.
In this case, the produced magnetic field would have a highly tangled configuration.
In the context of present-day star formation, numerical simulations demonstrated that 
such a tangled field tends to reduce magnetic braking efficiency and leads to form a larger disc 
(e.g., \citealp{Seifried2012}; \citealp{Joos2013}; \citealp{Tsukamoto2016}).
In addition, when the fields are completely tangled without any uniform field component, 
outflows strong enough to affect the stellar mass is hard to be launched (\citealp{Gerrard2019}).
Therefore, we need to investigate effects of field configuration on first-star formation in future works.\par
In our simulations, we have assumed ideal MHD since the field dissipation in the primordial gas is ineffective due to higher ionization degree than in the present-day case (\citealp{Maki_Susa2004}, \citeyear{Maki_Susa2007}; \citealp{Nakauchi2019}).
However, if the field is amplified to the critical field strength by a turbulence or a rotational motion in the accretion phase,
the ambipolar diffusion may become important in the primordial gas (\citealp{Schleicher2009}; \citealp{Nakauchi2019}).
The non-ideal effects have been well investigated in the present-day star formation (e.g., \citealp{Machida2006}; \citealp{Tomida2015}; \citealp{Vaytet2018}),
and it is known that the dissipation processes affect the disc and binary evolution by reducing the magnetic braking.
Therefore, future study for accretion phase of first star formation should also consider this dissipation effect.\par
The effects of MHD outflows during the accretion phase have already been studied for the present-day star formation with high accretion rate, $\dot{M}\sim 10^{-3}-10^{-2}\ M_{\odot}\ \rm yr^{-1}$ (e.g., \citealp{Matsushita2017}; \citealp{Machida2020}).
They have suggested that a massive outflow appears only when the initial cloud is strongly magnetized, i.e., $E_{\rm m}/|E_{\rm g}| \geq 0.1$.
On the other hand, with weaker fields, the outflows, even if launched, are weakened by ram pressure of the accretion flow in the course of propagation in the cloud (\citealp{Machida2020}).
Their results indicate that all outflows seen in our simulations do not necessarily grow enough to affect the star formation efficiency.
However, we cannot directly apply their results to the first star formation 
because the outflow speeds and launching scales are largely different from the present-day star formation due to the different thermal evolution.
In addition, the outflows are also launched from the first core scale in the present-day case.\par
For the first star formation, the ionization feedback from a protostar also has a significant effect on limiting its growth 
(e.g., \citealp{McKee_Tan2008}; \citealp{Hosokawa2011}).
Both the power of the ionization feedback and the MHD outflow depend on the strength of the accretion rate, 
and there would be mutual interaction between these two effects.
Therefore, in the future, we need to perform radiative MHD simulation taking into account both those effects in the accretion phase 
to reveal the effect of magnetic fields on the nature of the first stars.
\label{Sec:discussion_conclusion}
%%%%%%%%%%%%%%%%%%%%%%%%%%%%%%%%%%%%%%%%%%%%%%%%%%
%%%%%%%%%%%%%%%%%%%%%%%%%%%%%%%%%%%%%%%%%%%%%%%%%%
\section*{Acknowledgments}
The authors would like to thank Masahiro Machida and Sunmyon Chon for their helpful comments and discussion.
We also thank Daisuke Nakauchi and Ryoki Matsukoba for providing the modules for chemical network and cooling processes.  
KES acknowledges financial support from the Graduate Program on Physics for Universe of Tohoku University. 
KS appreciates the support by the Fellowship of the Japan Society for the Promotion of Science for Research Abroad.
The numerical simulations were performed on the Cray XC50 at CfCA of the National Astronomical Observatory of Japan, 
the computer cluster Draco at Frontier Research Institute for Interdisciplinary Sciences of Tohoku University,
and the Cray XC40 at Yukawa Institute for Theoretical Physics in Kyoto University.
This work was supported in part by the Grant-in-Aid from the Ministry of Education, Culture, Sports, Science and Technology (MEXT) of Japan 
(KO:17H02869, 17H01102, TM:18H05436, 18H05437, KT:16H05998, 18H05440).
%%%%%%%%%%%%%%%%%%%%%%%%%%%%%%%%%%%%%%%%%%%%%%%%%%
%%%%%%%%%%%%%%%%%%%%%%%%%%%%%%%%%%%%%%%%%%%%%%%%%%
\section*{Data Availability}
The data underlying this article will be shared on reasonable request to the corresponding author.

%%%%%%%%%%%%%%%%%%%% REFERENCES %%%%%%%%%%%%%%%%%%

% The best way to enter references is to use BibTeX:

\bibliographystyle{mnras}
\bibliography{hoge} % if your bibtex file is called example.bib

\begin{thebibliography}{}
\makeatletter
\relax
\def\mn@urlcharsother{\let\do\@makeother \do\$\do\&\do\#\do\^\do\_\do\%\do\~}
\def\mn@doi{\begingroup\mn@urlcharsother \@ifnextchar [ {\mn@doi@}
  {\mn@doi@[]}}
\def\mn@doi@[#1]#2{\def\@tempa{#1}\ifx\@tempa\@empty \href
  {http://dx.doi.org/#2} {doi:#2}\else \href {http://dx.doi.org/#2} {#1}\fi
  \endgroup}
\def\mn@eprint#1#2{\mn@eprint@#1:#2::\@nil}
\def\mn@eprint@arXiv#1{\href {http://arxiv.org/abs/#1} {{\tt arXiv:#1}}}
\def\mn@eprint@dblp#1{\href {http://dblp.uni-trier.de/rec/bibtex/#1.xml}
  {dblp:#1}}
\def\mn@eprint@#1:#2:#3:#4\@nil{\def\@tempa {#1}\def\@tempb {#2}\def\@tempc
  {#3}\ifx \@tempc \@empty \let \@tempc \@tempb \let \@tempb \@tempa \fi \ifx
  \@tempb \@empty \def\@tempb {arXiv}\fi \@ifundefined
  {mn@eprint@\@tempb}{\@tempb:\@tempc}{\expandafter \expandafter \csname
  mn@eprint@\@tempb\endcsname \expandafter{\@tempc}}}

\bibitem[\protect\citeauthoryear{{Abel}, {Bryan}  \& {Norman}}{{Abel}
  et~al.}{2002}]{Abel2002}
{Abel} T.,  {Bryan} G.~L.,   {Norman} M.~L.,  2002, \mn@doi [Science]
  {10.1126/science.295.5552.93}, \href
  {https://ui.adsabs.harvard.edu/abs/2002Sci...295...93A} {295, 93}

\bibitem[\protect\citeauthoryear{{Attia}, {Teyssier}, {Katz}, {Kimm},
  {Martin-Alvarez}, {Ocvirk}  \& {Rosdahl}}{{Attia} et~al.}{2021}]{Attia2021}
{Attia} O.,  {Teyssier} R.,  {Katz} H.,  {Kimm} T.,  {Martin-Alvarez} S.,
  {Ocvirk} P.,   {Rosdahl} J.,  2021, arXiv e-prints, \href
  {https://ui.adsabs.harvard.edu/abs/2021arXiv210209535A} {p. arXiv:2102.09535}

\bibitem[\protect\citeauthoryear{{Banerjee} \& {Jedamzik}}{{Banerjee} \&
  {Jedamzik}}{2004}]{Banerjee_Jedamzik2004}
{Banerjee} R.,  {Jedamzik} K.,  2004, \mn@doi [\prd]
  {10.1103/PhysRevD.70.123003}, \href
  {https://ui.adsabs.harvard.edu/abs/2004PhRvD..70l3003B} {70, 123003}

\bibitem[\protect\citeauthoryear{{Banerjee} \& {Pudritz}}{{Banerjee} \&
  {Pudritz}}{2006}]{Banerjee_Pudritz2006}
{Banerjee} R.,  {Pudritz} R.~E.,  2006, \mn@doi [\apj] {10.1086/500496}, \href
  {https://ui.adsabs.harvard.edu/abs/2006ApJ...641..949B} {641, 949}

\bibitem[\protect\citeauthoryear{{Baym}, {B{\"o}deker}  \& {McLerran}}{{Baym}
  et~al.}{1996}]{Baym1996}
{Baym} G.,  {B{\"o}deker} D.,   {McLerran} L.,  1996, \mn@doi [\prd]
  {10.1103/PhysRevD.53.662}, \href
  {https://ui.adsabs.harvard.edu/abs/1996PhRvD..53..662B} {53, 662}

\bibitem[\protect\citeauthoryear{{Biermann}}{{Biermann}}{1950}]{Biermann1950}
{Biermann} L.,  1950, Zeitschrift Naturforschung Teil A, \href
  {https://ui.adsabs.harvard.edu/abs/1950ZNatA...5...65B} {5, 65}

\bibitem[\protect\citeauthoryear{{Blandford} \& {Payne}}{{Blandford} \&
  {Payne}}{1982}]{Blandford_Payne1982}
{Blandford} R.~D.,  {Payne} D.~G.,  1982, \mn@doi [\mnras]
  {10.1093/mnras/199.4.883}, \href
  {https://ui.adsabs.harvard.edu/abs/1982MNRAS.199..883B} {199, 883}

\bibitem[\protect\citeauthoryear{{Bonnor}}{{Bonnor}}{1956}]{Bonnor1956}
{Bonnor} W.~B.,  1956, \mn@doi [\mnras] {10.1093/mnras/116.3.351}, \href
  {https://ui.adsabs.harvard.edu/abs/1956MNRAS.116..351B} {116, 351}

\bibitem[\protect\citeauthoryear{{Brandenburg} \& {Subramanian}}{{Brandenburg}
  \& {Subramanian}}{2005}]{Brandenburg_Subramanian2005}
{Brandenburg} A.,  {Subramanian} K.,  2005, \mn@doi [\physrep]
  {10.1016/j.physrep.2005.06.005}, \href
  {https://ui.adsabs.harvard.edu/abs/2005PhR...417....1B} {417, 1}

\bibitem[\protect\citeauthoryear{{Bromm}, {Coppi}  \& {Larson}}{{Bromm}
  et~al.}{2002}]{Bromm2002}
{Bromm} V.,  {Coppi} P.~S.,   {Larson} R.~B.,  2002, \mn@doi [\apj]
  {10.1086/323947}, \href
  {https://ui.adsabs.harvard.edu/abs/2002ApJ...564...23B} {564, 23}

\bibitem[\protect\citeauthoryear{{Chon} \& {Hosokawa}}{{Chon} \&
  {Hosokawa}}{2019}]{Chon_Hosokawa2019}
{Chon} S.,  {Hosokawa} T.,  2019, \mn@doi [\mnras] {10.1093/mnras/stz1824},
  \href {https://ui.adsabs.harvard.edu/abs/2019MNRAS.488.2658C} {488, 2658}

\bibitem[\protect\citeauthoryear{{Ciardi} \& {Ferrara}}{{Ciardi} \&
  {Ferrara}}{2005}]{Ciardi_Ferrara2005}
{Ciardi} B.,  {Ferrara} A.,  2005, \mn@doi [\ssr] {10.1007/s11214-005-3592-0},
  \href {https://ui.adsabs.harvard.edu/abs/2005SSRv..116..625C} {116, 625}

\bibitem[\protect\citeauthoryear{{Clark}, {Glover}, {Smith}, {Greif}, {Klessen}
   \& {Bromm}}{{Clark} et~al.}{2011}]{Clark2011}
{Clark} P.~C.,  {Glover} S. C.~O.,  {Smith} R.~J.,  {Greif} T.~H.,  {Klessen}
  R.~S.,   {Bromm} V.,  2011, \mn@doi [Science] {10.1126/science.1198027},
  \href {https://ui.adsabs.harvard.edu/abs/2011Sci...331.1040C} {331, 1040}

\bibitem[\protect\citeauthoryear{{Couchman} \& {Rees}}{{Couchman} \&
  {Rees}}{1986}]{Couchman_Rees1986}
{Couchman} H.~M.~P.,  {Rees} M.~J.,  1986, \mn@doi [\mnras]
  {10.1093/mnras/221.1.53}, \href
  {https://ui.adsabs.harvard.edu/abs/1986MNRAS.221...53C} {221, 53}

\bibitem[\protect\citeauthoryear{Dedner, Kemm, Kr^^c3^^b6ner, Munz, Schnitzer
  \& Wesenberg}{Dedner et~al.}{2002}]{Dedner2002}
Dedner A.,  Kemm F.,  Kr^^c3^^b6ner D.,  Munz C.-D.,  Schnitzer T.,   Wesenberg
  M.,  2002, \mn@doi [Journal of Computational Physics]
  {https://doi.org/10.1006/jcph.2001.6961}, 175, 645

\bibitem[\protect\citeauthoryear{{Doi} \& {Susa}}{{Doi} \&
  {Susa}}{2011}]{Doi_Susa2011}
{Doi} K.,  {Susa} H.,  2011, \mn@doi [\apj] {10.1088/0004-637X/741/2/93}, \href
  {https://ui.adsabs.harvard.edu/abs/2011ApJ...741...93D} {741, 93}

\bibitem[\protect\citeauthoryear{{Durrer} \& {Neronov}}{{Durrer} \&
  {Neronov}}{2013}]{Durrer_Neronov2013}
{Durrer} R.,  {Neronov} A.,  2013, \mn@doi [\aapr] {10.1007/s00159-013-0062-7},
  \href {https://ui.adsabs.harvard.edu/abs/2013A&ARv..21...62D} {21, 62}

\bibitem[\protect\citeauthoryear{{Ebert}}{{Ebert}}{1955}]{Ebert1955}
{Ebert} R.,  1955, \zap, \href
  {https://ui.adsabs.harvard.edu/abs/1955ZA.....37..217E} {37, 217}

\bibitem[\protect\citeauthoryear{{Federrath}, {Sur}, {Schleicher}, {Banerjee}
  \& {Klessen}}{{Federrath} et~al.}{2011}]{Federrath2011b}
{Federrath} C.,  {Sur} S.,  {Schleicher} D. R.~G.,  {Banerjee} R.,   {Klessen}
  R.~S.,  2011, \mn@doi [\apj] {10.1088/0004-637X/731/1/62}, \href
  {https://ui.adsabs.harvard.edu/abs/2011ApJ...731...62F} {731, 62}

\bibitem[\protect\citeauthoryear{{Fukushima}, {Omukai}  \&
  {Hosokawa}}{{Fukushima} et~al.}{2018}]{Fukushima2018}
{Fukushima} H.,  {Omukai} K.,   {Hosokawa} T.,  2018, \mn@doi [\mnras]
  {10.1093/mnras/stx2620}, \href
  {https://ui.adsabs.harvard.edu/abs/2018MNRAS.473.4754F} {473, 4754}

\bibitem[\protect\citeauthoryear{{Galli} \& {Shu}}{{Galli} \&
  {Shu}}{1993}]{Galli_Shu1993}
{Galli} D.,  {Shu} F.~H.,  1993, \mn@doi [\apj] {10.1086/173305}, \href
  {https://ui.adsabs.harvard.edu/abs/1993ApJ...417..220G} {417, 220}

\bibitem[\protect\citeauthoryear{{Gerrard}, {Federrath}  \&
  {Kuruwita}}{{Gerrard} et~al.}{2019}]{Gerrard2019}
{Gerrard} I.~A.,  {Federrath} C.,   {Kuruwita} R.,  2019, \mn@doi [\mnras]
  {10.1093/mnras/stz784}, \href
  {https://ui.adsabs.harvard.edu/abs/2019MNRAS.485.5532G} {485, 5532}

\bibitem[\protect\citeauthoryear{{Glover}}{{Glover}}{2015}]{Glover2015}
{Glover} S. C.~O.,  2015, \mn@doi [\mnras] {10.1093/mnras/stv1781}, \href
  {https://ui.adsabs.harvard.edu/abs/2015MNRAS.453.2901G} {453, 2901}

\bibitem[\protect\citeauthoryear{{Gnedin}, {Ferrara}  \& {Zweibel}}{{Gnedin}
  et~al.}{2000}]{Gnedin2000}
{Gnedin} N.~Y.,  {Ferrara} A.,   {Zweibel} E.~G.,  2000, \mn@doi [\apj]
  {10.1086/309272}, \href
  {https://ui.adsabs.harvard.edu/abs/2000ApJ...539..505G} {539, 505}

\bibitem[\protect\citeauthoryear{{Greif}}{{Greif}}{2015}]{Greif2015}
{Greif} T.~H.,  2015, \mn@doi [Computational Astrophysics and Cosmology]
  {10.1186/s40668-014-0006-2}, \href
  {https://ui.adsabs.harvard.edu/abs/2015ComAC...2....3G} {2, 3}

\bibitem[\protect\citeauthoryear{{Hanayama}, {Takahashi}, {Kotake}, {Oguri},
  {Ichiki}  \& {Ohno}}{{Hanayama} et~al.}{2005}]{Hanayama2005}
{Hanayama} H.,  {Takahashi} K.,  {Kotake} K.,  {Oguri} M.,  {Ichiki} K.,
  {Ohno} H.,  2005, \mn@doi [\apj] {10.1086/491575}, \href
  {https://ui.adsabs.harvard.edu/abs/2005ApJ...633..941H} {633, 941}

\bibitem[\protect\citeauthoryear{{Hirano}, {Hosokawa}, {Yoshida}, {Umeda},
  {Omukai}, {Chiaki}  \& {Yorke}}{{Hirano} et~al.}{2014}]{Hirano2014}
{Hirano} S.,  {Hosokawa} T.,  {Yoshida} N.,  {Umeda} H.,  {Omukai} K.,
  {Chiaki} G.,   {Yorke} H.~W.,  2014, \mn@doi [\apj]
  {10.1088/0004-637X/781/2/60}, \href
  {https://ui.adsabs.harvard.edu/abs/2014ApJ...781...60H} {781, 60}

\bibitem[\protect\citeauthoryear{{Hosokawa}, {Omukai}, {Yoshida}  \&
  {Yorke}}{{Hosokawa} et~al.}{2011}]{Hosokawa2011}
{Hosokawa} T.,  {Omukai} K.,  {Yoshida} N.,   {Yorke} H.~W.,  2011, \mn@doi
  [Science] {10.1126/science.1207433}, \href
  {https://ui.adsabs.harvard.edu/abs/2011Sci...334.1250H} {334, 1250}

\bibitem[\protect\citeauthoryear{{Hosokawa}, {Hirano}, {Kuiper}, {Yorke},
  {Omukai}  \& {Yoshida}}{{Hosokawa} et~al.}{2016}]{Hosokawa2016}
{Hosokawa} T.,  {Hirano} S.,  {Kuiper} R.,  {Yorke} H.~W.,  {Omukai} K.,
  {Yoshida} N.,  2016, \mn@doi [\apj] {10.3847/0004-637X/824/2/119}, \href
  {https://ui.adsabs.harvard.edu/abs/2016ApJ...824..119H} {824, 119}

\bibitem[\protect\citeauthoryear{{Ichiki}, {Takahashi}, {Ohno}, {Hanayama}  \&
  {Sugiyama}}{{Ichiki} et~al.}{2006}]{Ichiki2006}
{Ichiki} K.,  {Takahashi} K.,  {Ohno} H.,  {Hanayama} H.,   {Sugiyama} N.,
  2006, \mn@doi [Science] {10.1126/science.1120690}, \href
  {https://ui.adsabs.harvard.edu/abs/2006Sci...311..827I} {311, 827}

\bibitem[\protect\citeauthoryear{{Joos}, {Hennebelle}, {Ciardi}  \&
  {Fromang}}{{Joos} et~al.}{2013}]{Joos2013}
{Joos} M.,  {Hennebelle} P.,  {Ciardi} A.,   {Fromang} S.,  2013, \mn@doi
  [\aap] {10.1051/0004-6361/201220649}, \href
  {https://ui.adsabs.harvard.edu/abs/2013A&A...554A..17J} {554, A17}

\bibitem[\protect\citeauthoryear{{Kimura}, {Hosokawa}  \& {Sugimura}}{{Kimura}
  et~al.}{2020}]{Kimura2020}
{Kimura} K.,  {Hosokawa} T.,   {Sugimura} K.,  2020, arXiv e-prints, \href
  {https://ui.adsabs.harvard.edu/abs/2020arXiv201201452K} {p. arXiv:2012.01452}

\bibitem[\protect\citeauthoryear{{Kinugawa}, {Inayoshi}, {Hotokezaka},
  {Nakauchi}  \& {Nakamura}}{{Kinugawa} et~al.}{2014}]{Kinugawa2014}
{Kinugawa} T.,  {Inayoshi} K.,  {Hotokezaka} K.,  {Nakauchi} D.,   {Nakamura}
  T.,  2014, \mn@doi [\mnras] {10.1093/mnras/stu1022}, \href
  {https://ui.adsabs.harvard.edu/abs/2014MNRAS.442.2963K} {442, 2963}

\bibitem[\protect\citeauthoryear{{Kinugawa}, {Miyamoto}, {Kanda}  \&
  {Nakamura}}{{Kinugawa} et~al.}{2016}]{Kinugawa2016}
{Kinugawa} T.,  {Miyamoto} A.,  {Kanda} N.,   {Nakamura} T.,  2016, \mn@doi
  [\mnras] {10.1093/mnras/stv2624}, \href
  {https://ui.adsabs.harvard.edu/abs/2016MNRAS.456.1093K} {456, 1093}

\bibitem[\protect\citeauthoryear{{Kulsrud}, {Cen}, {Ostriker}  \&
  {Ryu}}{{Kulsrud} et~al.}{1997}]{Kulsrud1997}
{Kulsrud} R.~M.,  {Cen} R.,  {Ostriker} J.~P.,   {Ryu} D.,  1997, \mn@doi
  [\apj] {10.1086/303987}, \href
  {https://ui.adsabs.harvard.edu/abs/1997ApJ...480..481K} {480, 481}

\bibitem[\protect\citeauthoryear{{Langer}, {Puget}  \& {Aghanim}}{{Langer}
  et~al.}{2003}]{Langer2003}
{Langer} M.,  {Puget} J.-L.,   {Aghanim} N.,  2003, \mn@doi [\prd]
  {10.1103/PhysRevD.67.043505}, \href
  {https://ui.adsabs.harvard.edu/abs/2003PhRvD..67d3505L} {67, 043505}

\bibitem[\protect\citeauthoryear{{Larson}}{{Larson}}{1969}]{Larson1969}
{Larson} R.~B.,  1969, \mn@doi [\mnras] {10.1093/mnras/145.3.271}, \href
  {https://ui.adsabs.harvard.edu/abs/1969MNRAS.145..271L} {145, 271}

\bibitem[\protect\citeauthoryear{{Lipovka}, {N{\'u}{\~n}ez-L{\'o}pez}  \&
  {Avila-Reese}}{{Lipovka} et~al.}{2005}]{Lipovka2005}
{Lipovka} A.,  {N{\'u}{\~n}ez-L{\'o}pez} R.,   {Avila-Reese} V.,  2005, \mn@doi
  [\mnras] {10.1111/j.1365-2966.2005.09226.x}, \href
  {https://ui.adsabs.harvard.edu/abs/2005MNRAS.361..850L} {361, 850}

\bibitem[\protect\citeauthoryear{{Machida} \& {Hosokawa}}{{Machida} \&
  {Hosokawa}}{2020}]{Machida2020}
{Machida} M.~N.,  {Hosokawa} T.,  2020, \mn@doi [\mnras]
  {10.1093/mnras/staa3139}, \href
  {https://ui.adsabs.harvard.edu/abs/2020MNRAS.499.4490M} {499, 4490}

\bibitem[\protect\citeauthoryear{{Machida}, {Inutsuka}  \&
  {Matsumoto}}{{Machida} et~al.}{2006}]{Machida2006}
{Machida} M.~N.,  {Inutsuka} S.-i.,   {Matsumoto} T.,  2006, \mn@doi [\apjl]
  {10.1086/507179}, \href
  {https://ui.adsabs.harvard.edu/abs/2006ApJ...647L.151M} {647, L151}

\bibitem[\protect\citeauthoryear{{Machida}, {Inutsuka}  \&
  {Matsumoto}}{{Machida} et~al.}{2008a}]{Machida2008a}
{Machida} M.~N.,  {Inutsuka} S.-i.,   {Matsumoto} T.,  2008a, \mn@doi [\apj]
  {10.1086/528364}, \href
  {https://ui.adsabs.harvard.edu/abs/2008ApJ...676.1088M} {676, 1088}

\bibitem[\protect\citeauthoryear{{Machida}, {Omukai}, {Matsumoto}  \&
  {Inutsuka}}{{Machida} et~al.}{2008b}]{Machida2008b}
{Machida} M.~N.,  {Omukai} K.,  {Matsumoto} T.,   {Inutsuka} S.-i.,  2008b,
  \mn@doi [\apj] {10.1086/533434}, \href
  {https://ui.adsabs.harvard.edu/abs/2008ApJ...677..813M} {677, 813}

\bibitem[\protect\citeauthoryear{{Machida}, {Matsumoto}  \&
  {Inutsuka}}{{Machida} et~al.}{2008c}]{Machida2008}
{Machida} M.~N.,  {Matsumoto} T.,   {Inutsuka} S.-i.,  2008c, \mn@doi [\apj]
  {10.1086/591074}, \href
  {https://ui.adsabs.harvard.edu/abs/2008ApJ...685..690M} {685, 690}

\bibitem[\protect\citeauthoryear{{Maki} \& {Susa}}{{Maki} \&
  {Susa}}{2004}]{Maki_Susa2004}
{Maki} H.,  {Susa} H.,  2004, \mn@doi [\apj] {10.1086/421103}, \href
  {https://ui.adsabs.harvard.edu/abs/2004ApJ...609..467M} {609, 467}

\bibitem[\protect\citeauthoryear{{Maki} \& {Susa}}{{Maki} \&
  {Susa}}{2007}]{Maki_Susa2007}
{Maki} H.,  {Susa} H.,  2007, \mn@doi [\pasj] {10.1093/pasj/59.4.787}, \href
  {https://ui.adsabs.harvard.edu/abs/2007PASJ...59..787M} {59, 787}

\bibitem[\protect\citeauthoryear{{Marchand}, {Tomida}, {Tanaka},
  {Commer{\c{c}}on}  \& {Chabrier}}{{Marchand} et~al.}{2020}]{Marchand2020}
{Marchand} P.,  {Tomida} K.,  {Tanaka} K. E.~I.,  {Commer{\c{c}}on} B.,
  {Chabrier} G.,  2020, \mn@doi [\apj] {10.3847/1538-4357/abad99}, \href
  {https://ui.adsabs.harvard.edu/abs/2020ApJ...900..180M} {900, 180}

\bibitem[\protect\citeauthoryear{{Matsukoba}, {Takahashi}, {Sugimura}  \&
  {Omukai}}{{Matsukoba} et~al.}{2019}]{Matsukoba2019}
{Matsukoba} R.,  {Takahashi} S.~Z.,  {Sugimura} K.,   {Omukai} K.,  2019,
  \mn@doi [\mnras] {10.1093/mnras/sty3522}, \href
  {https://ui.adsabs.harvard.edu/abs/2019MNRAS.484.2605M} {484, 2605}

\bibitem[\protect\citeauthoryear{{Matsumoto}}{{Matsumoto}}{2007}]{Matsumoto2007}
{Matsumoto} T.,  2007, \mn@doi [\pasj] {10.1093/pasj/59.5.905}, \href
  {https://ui.adsabs.harvard.edu/abs/2007PASJ...59..905M} {59, 905}

\bibitem[\protect\citeauthoryear{{Matsumoto} \& {Tomisaka}}{{Matsumoto} \&
  {Tomisaka}}{2004}]{Matsumoto_Tomisaka2004}
{Matsumoto} T.,  {Tomisaka} K.,  2004, \mn@doi [\apj] {10.1086/424897}, \href
  {https://ui.adsabs.harvard.edu/abs/2004ApJ...616..266M} {616, 266}

\bibitem[\protect\citeauthoryear{{Matsumoto}, {Dobashi}  \&
  {Shimoikura}}{{Matsumoto} et~al.}{2015}]{Matsumoto2015}
{Matsumoto} T.,  {Dobashi} K.,   {Shimoikura} T.,  2015, \mn@doi [\apj]
  {10.1088/0004-637X/801/2/77}, \href
  {https://ui.adsabs.harvard.edu/abs/2015ApJ...801...77M} {801, 77}

\bibitem[\protect\citeauthoryear{{Matsushita}, {Machida}, {Sakurai}  \&
  {Hosokawa}}{{Matsushita} et~al.}{2017}]{Matsushita2017}
{Matsushita} Y.,  {Machida} M.~N.,  {Sakurai} Y.,   {Hosokawa} T.,  2017,
  \mn@doi [\mnras] {10.1093/mnras/stx893}, \href
  {https://ui.adsabs.harvard.edu/abs/2017MNRAS.470.1026M} {470, 1026}

\bibitem[\protect\citeauthoryear{{McKee} \& {Tan}}{{McKee} \&
  {Tan}}{2008}]{McKee_Tan2008}
{McKee} C.~F.,  {Tan} J.~C.,  2008, \mn@doi [\apj] {10.1086/587434}, \href
  {https://ui.adsabs.harvard.edu/abs/2008ApJ...681..771M} {681, 771}

\bibitem[\protect\citeauthoryear{{Miyoshi} \& {Kusano}}{{Miyoshi} \&
  {Kusano}}{2005}]{Miyoshi_Kusano2005}
{Miyoshi} T.,  {Kusano} K.,  2005, in AGU Fall Meeting Abstracts. pp
  SM51B--1295

\bibitem[\protect\citeauthoryear{{Mouschovias} \& {Paleologou}}{{Mouschovias}
  \& {Paleologou}}{1979}]{Mouschovias_Paleologou1979}
{Mouschovias} T.~C.,  {Paleologou} E.~V.,  1979, \mn@doi [\apj]
  {10.1086/157077}, \href
  {https://ui.adsabs.harvard.edu/abs/1979ApJ...230..204M} {230, 204}

\bibitem[\protect\citeauthoryear{{Mouschovias} \& {Spitzer}}{{Mouschovias} \&
  {Spitzer}}{1976}]{Mouschovias_Spitzer1976}
{Mouschovias} T.~C.,  {Spitzer} L. J.,  1976, \mn@doi [\apj] {10.1086/154835},
  \href {https://ui.adsabs.harvard.edu/abs/1976ApJ...210..326M} {210, 326}

\bibitem[\protect\citeauthoryear{{Nakauchi}, {Omukai}  \& {Susa}}{{Nakauchi}
  et~al.}{2019}]{Nakauchi2019}
{Nakauchi} D.,  {Omukai} K.,   {Susa} H.,  2019, \mn@doi [\mnras]
  {10.1093/mnras/stz1799}, \href
  {https://ui.adsabs.harvard.edu/abs/2019MNRAS.488.1846N} {488, 1846}

\bibitem[\protect\citeauthoryear{{Omukai} \& {Nishi}}{{Omukai} \&
  {Nishi}}{1998}]{Omukai_Nishi1998}
{Omukai} K.,  {Nishi} R.,  1998, \mn@doi [\apj] {10.1086/306395}, \href
  {https://ui.adsabs.harvard.edu/abs/1998ApJ...508..141O} {508, 141}

\bibitem[\protect\citeauthoryear{{Omukai} \& {Palla}}{{Omukai} \&
  {Palla}}{2003}]{Omukai_Palla2003}
{Omukai} K.,  {Palla} F.,  2003, \mn@doi [\apj] {10.1086/374810}, \href
  {https://ui.adsabs.harvard.edu/abs/2003ApJ...589..677O} {589, 677}

\bibitem[\protect\citeauthoryear{{Omukai}, {Tsuribe}, {Schneider}  \&
  {Ferrara}}{{Omukai} et~al.}{2005}]{Omukai2005}
{Omukai} K.,  {Tsuribe} T.,  {Schneider} R.,   {Ferrara} A.,  2005, \mn@doi
  [\apj] {10.1086/429955}, \href
  {https://ui.adsabs.harvard.edu/abs/2005ApJ...626..627O} {626, 627}

\bibitem[\protect\citeauthoryear{{Omukai}, {Hosokawa}  \& {Yoshida}}{{Omukai}
  et~al.}{2010}]{Omukai2010}
{Omukai} K.,  {Hosokawa} T.,   {Yoshida} N.,  2010, \mn@doi [\apj]
  {10.1088/0004-637X/722/2/1793}, \href
  {https://ui.adsabs.harvard.edu/abs/2010ApJ...722.1793O} {722, 1793}

\bibitem[\protect\citeauthoryear{{Pakmor} et~al.,}{{Pakmor}
  et~al.}{2017}]{Pakmor2017}
{Pakmor} R.,  et~al., 2017, \mn@doi [\mnras] {10.1093/mnras/stx1074}, \href
  {https://ui.adsabs.harvard.edu/abs/2017MNRAS.469.3185P} {469, 3185}

\bibitem[\protect\citeauthoryear{{Penston}}{{Penston}}{1969}]{Penston1969}
{Penston} M.~V.,  1969, \mn@doi [\mnras] {10.1093/mnras/144.4.425}, \href
  {https://ui.adsabs.harvard.edu/abs/1969MNRAS.144..425P} {144, 425}

\bibitem[\protect\citeauthoryear{{Quashnock}, {Loeb}  \& {Spergel}}{{Quashnock}
  et~al.}{1989}]{Quashnock1989}
{Quashnock} J.~M.,  {Loeb} A.,   {Spergel} D.~N.,  1989, \mn@doi [\apjl]
  {10.1086/185528}, \href
  {https://ui.adsabs.harvard.edu/abs/1989ApJ...344L..49Q} {344, L49}

\bibitem[\protect\citeauthoryear{{Schleicher}, {Galli}, {Glover}, {Banerjee},
  {Palla}, {Schneider}  \& {Klessen}}{{Schleicher}
  et~al.}{2009}]{Schleicher2009}
{Schleicher} D. R.~G.,  {Galli} D.,  {Glover} S. C.~O.,  {Banerjee} R.,
  {Palla} F.,  {Schneider} R.,   {Klessen} R.~S.,  2009, \mn@doi [\apj]
  {10.1088/0004-637X/703/1/1096}, \href
  {https://ui.adsabs.harvard.edu/abs/2009ApJ...703.1096S} {703, 1096}

\bibitem[\protect\citeauthoryear{{Schober}, {Schleicher}, {Federrath},
  {Glover}, {Klessen}  \& {Banerjee}}{{Schober} et~al.}{2012}]{Schober2012}
{Schober} J.,  {Schleicher} D.,  {Federrath} C.,  {Glover} S.,  {Klessen}
  R.~S.,   {Banerjee} R.,  2012, \mn@doi [\apj] {10.1088/0004-637X/754/2/99},
  \href {https://ui.adsabs.harvard.edu/abs/2012ApJ...754...99S} {754, 99}

\bibitem[\protect\citeauthoryear{{Seifried}, {Banerjee}, {Pudritz}  \&
  {Klessen}}{{Seifried} et~al.}{2012}]{Seifried2012}
{Seifried} D.,  {Banerjee} R.,  {Pudritz} R.~E.,   {Klessen} R.~S.,  2012,
  \mn@doi [\mnras] {10.1111/j.1745-3933.2012.01253.x}, \href
  {https://ui.adsabs.harvard.edu/abs/2012MNRAS.423L..40S} {423, L40}

\bibitem[\protect\citeauthoryear{{Sharda}, {Federrath}, {Krumholz}  \&
  {Schleicher}}{{Sharda} et~al.}{2020a}]{Sharda2020b}
{Sharda} P.,  {Federrath} C.,  {Krumholz} M.~R.,   {Schleicher} D. R.~G.,
  2020a, arXiv e-prints, \href
  {https://ui.adsabs.harvard.edu/abs/2020arXiv200702678S} {p. arXiv:2007.02678}

\bibitem[\protect\citeauthoryear{{Sharda}, {Federrath}  \& {Krumholz}}{{Sharda}
  et~al.}{2020b}]{Sharda2020a}
{Sharda} P.,  {Federrath} C.,   {Krumholz} M.~R.,  2020b, \mn@doi [\mnras]
  {10.1093/mnras/staa1926}, \href
  {https://ui.adsabs.harvard.edu/abs/2020MNRAS.497..336S} {497, 336}

\bibitem[\protect\citeauthoryear{{Smith}, {Glover}, {Clark}, {Greif}  \&
  {Klessen}}{{Smith} et~al.}{2011}]{Smith2011}
{Smith} R.~J.,  {Glover} S. C.~O.,  {Clark} P.~C.,  {Greif} T.,   {Klessen}
  R.~S.,  2011, \mn@doi [\mnras] {10.1111/j.1365-2966.2011.18659.x}, \href
  {https://ui.adsabs.harvard.edu/abs/2011MNRAS.414.3633S} {414, 3633}

\bibitem[\protect\citeauthoryear{{Stacy}, {Bromm}  \& {Lee}}{{Stacy}
  et~al.}{2016}]{Stacy2016}
{Stacy} A.,  {Bromm} V.,   {Lee} A.~T.,  2016, \mn@doi [\mnras]
  {10.1093/mnras/stw1728}, \href
  {https://ui.adsabs.harvard.edu/abs/2016MNRAS.462.1307S} {462, 1307}

\bibitem[\protect\citeauthoryear{{Stahler}, {Palla}  \& {Salpeter}}{{Stahler}
  et~al.}{1986}]{Stahler_Palla1986}
{Stahler} S.~W.,  {Palla} F.,   {Salpeter} E.~E.,  1986, \mn@doi [\apj]
  {10.1086/164018}, \href
  {https://ui.adsabs.harvard.edu/abs/1986ApJ...302..590S} {302, 590}

\bibitem[\protect\citeauthoryear{{Subramanian}}{{Subramanian}}{2016}]{Subramanian2016}
{Subramanian} K.,  2016, \mn@doi [Reports on Progress in Physics]
  {10.1088/0034-4885/79/7/076901}, \href
  {https://ui.adsabs.harvard.edu/abs/2016RPPh...79g6901S} {79, 076901}

\bibitem[\protect\citeauthoryear{{Sugimura}, {Matsumoto}, {Hosokawa}, {Hirano}
  \& {Omukai}}{{Sugimura} et~al.}{2020}]{Sugimura2020}
{Sugimura} K.,  {Matsumoto} T.,  {Hosokawa} T.,  {Hirano} S.,   {Omukai} K.,
  2020, \mn@doi [\apjl] {10.3847/2041-8213/ab7d37}, \href
  {https://ui.adsabs.harvard.edu/abs/2020ApJ...892L..14S} {892, L14}

\bibitem[\protect\citeauthoryear{{Sur}, {Schleicher}, {Banerjee}, {Federrath}
  \& {Klessen}}{{Sur} et~al.}{2010}]{Sur2010}
{Sur} S.,  {Schleicher} D.~R.~G.,  {Banerjee} R.,  {Federrath} C.,   {Klessen}
  R.~S.,  2010, \mn@doi [\apjl] {10.1088/2041-8205/721/2/L134}, \href
  {https://ui.adsabs.harvard.edu/abs/2010ApJ...721L.134S} {721, L134}

\bibitem[\protect\citeauthoryear{{Susa}}{{Susa}}{2019}]{Susa2019}
{Susa} H.,  2019, \mn@doi [\apj] {10.3847/1538-4357/ab1b6f}, \href
  {https://ui.adsabs.harvard.edu/abs/2019ApJ...877...99S} {877, 99}

\bibitem[\protect\citeauthoryear{{Susa}, {Hasegawa}  \& {Tominaga}}{{Susa}
  et~al.}{2014}]{Susa2014}
{Susa} H.,  {Hasegawa} K.,   {Tominaga} N.,  2014, \mn@doi [\apj]
  {10.1088/0004-637X/792/1/32}, \href
  {https://ui.adsabs.harvard.edu/abs/2014ApJ...792...32S} {792, 32}

\bibitem[\protect\citeauthoryear{{Tanaka} \& {Omukai}}{{Tanaka} \&
  {Omukai}}{2014}]{Tanaka_Omukai2014}
{Tanaka} K. E.~I.,  {Omukai} K.,  2014, \mn@doi [\mnras]
  {10.1093/mnras/stu069}, \href
  {https://ui.adsabs.harvard.edu/abs/2014MNRAS.439.1884T} {439, 1884}

\bibitem[\protect\citeauthoryear{{Tomida}, {Tomisaka}, {Matsumoto}, {Hori},
  {Okuzumi}, {Machida}  \& {Saigo}}{{Tomida} et~al.}{2013}]{Tomida2013}
{Tomida} K.,  {Tomisaka} K.,  {Matsumoto} T.,  {Hori} Y.,  {Okuzumi} S.,
  {Machida} M.~N.,   {Saigo} K.,  2013, \mn@doi [\apj]
  {10.1088/0004-637X/763/1/6}, \href
  {https://ui.adsabs.harvard.edu/abs/2013ApJ...763....6T} {763, 6}

\bibitem[\protect\citeauthoryear{{Tomida}, {Okuzumi}  \& {Machida}}{{Tomida}
  et~al.}{2015}]{Tomida2015}
{Tomida} K.,  {Okuzumi} S.,   {Machida} M.~N.,  2015, \mn@doi [\apj]
  {10.1088/0004-637X/801/2/117}, \href
  {https://ui.adsabs.harvard.edu/abs/2015ApJ...801..117T} {801, 117}

\bibitem[\protect\citeauthoryear{{Tomisaka}}{{Tomisaka}}{2002}]{Tomisaka2002}
{Tomisaka} K.,  2002, \mn@doi [\apj] {10.1086/341133}, \href
  {https://ui.adsabs.harvard.edu/abs/2002ApJ...575..306T} {575, 306}

\bibitem[\protect\citeauthoryear{{Tsukamoto}}{{Tsukamoto}}{2016}]{Tsukamoto2016}
{Tsukamoto} Y.,  2016, \mn@doi [\pasa] {10.1017/pasa.2016.6}, \href
  {https://ui.adsabs.harvard.edu/abs/2016PASA...33...10T} {33, e010}

\bibitem[\protect\citeauthoryear{{Turk}, {Oishi}, {Abel}  \& {Bryan}}{{Turk}
  et~al.}{2012}]{Turk2012}
{Turk} M.~J.,  {Oishi} J.~S.,  {Abel} T.,   {Bryan} G.~L.,  2012, \mn@doi
  [\apj] {10.1088/0004-637X/745/2/154}, \href
  {https://ui.adsabs.harvard.edu/abs/2012ApJ...745..154T} {745, 154}

\bibitem[\protect\citeauthoryear{{Vaytet}, {Commer{\c{c}}on}, {Masson},
  {Gonz{\'a}lez}  \& {Chabrier}}{{Vaytet} et~al.}{2018}]{Vaytet2018}
{Vaytet} N.,  {Commer{\c{c}}on} B.,  {Masson} J.,  {Gonz{\'a}lez} M.,
  {Chabrier} G.,  2018, \mn@doi [\aap] {10.1051/0004-6361/201732075}, \href
  {https://ui.adsabs.harvard.edu/abs/2018A&A...615A...5V} {615, A5}

\bibitem[\protect\citeauthoryear{{Xu}, {O'Shea}, {Collins}, {Norman}, {Li}  \&
  {Li}}{{Xu} et~al.}{2008}]{Xu2008}
{Xu} H.,  {O'Shea} B.~W.,  {Collins} D.~C.,  {Norman} M.~L.,  {Li} H.,   {Li}
  S.,  2008, \mn@doi [\apjl] {10.1086/595617}, \href
  {https://ui.adsabs.harvard.edu/abs/2008ApJ...688L..57X} {688, L57}

\bibitem[\protect\citeauthoryear{{Yahil}}{{Yahil}}{1983}]{Yahil1983}
{Yahil} A.,  1983, \mn@doi [\apj] {10.1086/160746}, \href
  {https://ui.adsabs.harvard.edu/abs/1983ApJ...265.1047Y} {265, 1047}

\bibitem[\protect\citeauthoryear{{Yoshida}, {Abel}, {Hernquist}  \&
  {Sugiyama}}{{Yoshida} et~al.}{2003}]{Yoshida2003}
{Yoshida} N.,  {Abel} T.,  {Hernquist} L.,   {Sugiyama} N.,  2003, \mn@doi
  [\apj] {10.1086/375810}, \href
  {https://ui.adsabs.harvard.edu/abs/2003ApJ...592..645Y} {592, 645}

\bibitem[\protect\citeauthoryear{{Yoshida}, {Omukai}  \& {Hernquist}}{{Yoshida}
  et~al.}{2008}]{Yoshida2008}
{Yoshida} N.,  {Omukai} K.,   {Hernquist} L.,  2008, \mn@doi [Science]
  {10.1126/science.1160259}, \href
  {https://ui.adsabs.harvard.edu/abs/2008Sci...321..669Y} {321, 669}

\makeatother
\end{thebibliography}

% This method is tedious and prone to error if you have lots of references
%\begin{thebibliography}{99}
%\bibitem[\protect\citeauthoryear{Author}{2012}]{Author2012}
%Author A.~N., 2013, Journal of Improbable Astronomy, 1, 1
%\bibitem[\protect\citeauthoryear{Others}{2013}]{Others2013}
%Others S., 2012, Journal of Interesting Stuff, 17, 198
%\end{thebibliography}

%%%%%%%%%%%%%%%%%%%%%%%%%%%%%%%%%%%%%%%%%%%%%%%%%%

%%%%%%%%%%%%%%%%% APPENDICES %%%%%%%%%%%%%%%%%%%%%

%%%%%%%%%%%%%%%%%%%%%%%%%%%%%%%%%%%%%%%%%%%%%%%%%%

% Don't change these lines
\bsp	% typesetting comment
\label{lastpage}
\end{document}